\documentclass[useAMS,usenatbib]{mn2e}
\usepackage{graphicx}
\usepackage{lscape}
\usepackage{latexsym}
\usepackage{xspace}
\usepackage{rotating}
\usepackage{lscape}
\usepackage[percent]{overpic}
\usepackage{color}

 \title[]{On the nature of star-forming filaments: II. Sub-filaments and velocities}
\author[Smith et~al.]{Rowan J. Smith$^{1}$\thanks{Email: rowan.smith@manchester.ac.uk}, Simon C. O. Glover$^{2}$, Ralf. S. Klessen$^{2}$, Gary A. Fuller$^{1}$ \\
$^{1}$Jodrell Bank Centre for Astrophysics, School of Physics and Astronomy, University of Manchester, Oxford Road,Manchester, M13 9PL,UK\\
$^{2}$Zentrum f\"ur Astronomie der Universit\"at Heidelberg, Institut f\"ur Theoretische Astrophysik, Albert-Ueberle-Str. 2, 69120 Heidelberg, Germany \\
 }
 
\begin{document}

\pagerange{\pageref{firstpage}--\pageref{lastpage}} \pubyear{2015}

\maketitle

\label{firstpage}

\def\mnras{MNRAS}
\def\apj{ApJ}
\def\aap{A\&A}
\def\apjl{ApJL}
\def\apjs{ApJS}
\def\bain{BAIN}
\def\pasp{PASP}
\def\araa{ARA\&A}
\def\ga{\sim}
\def\nat{Nature}
\def\aj{AJ}
\def\pasj{PASJ}
\def\sovast{Soviet Astronomy}

%some simplifying definitions

\newcommand{\eq}{Equation }
\newcommand{\fig}{Figure }
\newcommand{\msun}{\,M$_{\odot}$ }
\newcommand{\gcmc}{\,g\,cm$^{-3}$}
\newcommand{\cmc}{\,cm$^{-3}$}
\newcommand{\cms}{\,cm$^{-2}$}
\newcommand{\kms}{\,km\,s$^{-1}$}
\newcommand{\tab}{Table }
\newcommand{\gcms}{\,g\,cm$^{-2}$\xspace}
\newcommand{\E}{\times 10}

% acronyms
\newcommand{\arepo}{\textsc{arepo}\xspace}
\newcommand{\sph}{\textsc{sph}\xspace}
\newcommand{\treecol}{\textsc{treecol}\xspace}
\newcommand{\healpix}{\textsc{healpix}\xspace}
\newcommand{\gadget}{\textsc{gadget} 2\xspace}
\newcommand{\disperse}{DisPerSE\xspace}
\newcommand{\mpfit}{\textsc{mpfit}\xspace}
\newcommand{\herschel}{\textit{Herschel}\xspace}
\newcommand{\radmc}{\textsc{radmc-3d}\xspace}

%  terminology
\newcommand{\vflow}{\textsc{$v_{\rm flow}$}\xspace}
\newcommand{\vfront}{\textsc{$v_{\rm front}$}\xspace}
\newcommand{\vrad}{\textsc{$v_{\rm rad}$}\xspace}
\newcommand{\vrot}{\textsc{$v_{\rm rot}$}\xspace}

\begin{abstract}
We show that hydrodynamic turbulent cloud simulations naturally produce large filaments made up of a network of smaller and coherent sub-filaments. Such simulations resemble observations of filaments and fibres in nearby molecular clouds. The sub-filaments are dynamical features formed at the stagnation points of the turbulent velocity field where shocks dissipate the turbulent energy. They are a ubiquitous feature of the simulated clouds, which appear from the beginning of the simulation and are not formed by gradual fragmentation of  larger filaments. Most of the sub-filaments are gravitationally sub-critical and do not fragment into cores, however, there is also a significant fraction of supercritical sub-filaments which break up into star-forming cores. The sub-filaments are coherent along their length, and the residual velocities along their spine show that they are subsonically contracting without any ordered rotation on scales of $\sim0.1$ pc. Accretion flows along the sub-filaments can feed material into star forming cores embedded within the network. The overall mass in sub-filaments and the number of sub-filaments increases as the cloud evolves. We propose that the formation of filaments and sub-filaments is a natural consequence of the turbulent cascade in the complex multi-phase interstellar medium. Sub-filaments are formed by the high wavenumber, small scale modes in the turbulent velocity field. These are then stretched by local shear motions and gathered together by a combination of low wavenumber modes and gravitational contraction on larger scales, and by doing so build up the extended filaments observed in column density maps.
\end{abstract}
\begin{keywords}
star formation, filaments, molecular clouds 
\end{keywords}
%Mass can flow along the sub-filaments and either increase or decrease their density. In particular, accretion flows along the sub-filaments can feed material into star forming cores embedded within the network.

\section{Introduction}\label{sec-intro}
%In which we motivate our study from the literature and present our main aims.

It has become clear that star-forming gas cores are not randomly distributed in molecular clouds, but instead are threaded along networks of filaments like beads on a string. This conclusion became inescapable following the observations of \herschel \citep{Andre10, Menshchikov10, Arzoumanian11, Hennemann12, Schneider12, Palmeirim13}, but had also been suggested in previous works \citep{Schneider79}, and in work with other instruments \citep{Schneider10,Juvela12a,Malinen12,Seo15}. Several authors have shown that filamentary flows may play a key role in star formation. For instance, \citet{Kirk13} show that in Serpens South there are mass flows of $3.0\E^{-5}$ \msun yr$^{-1}$ along the filament's long axis towards the central cluster, and radial mass flows onto the filament of $1.3\E^{-4}$ \msun yr$^{-1}$. Similarly, \citet{Peretto13} found an inflow rate of $2.5\E^{-3}$ \msun yr$^{-1}$ along filaments towards the central star forming clump in SDC335.579-0.272. In numerical simulations, cores embedded within filaments have been shown to form more massive stars  by accretion along filamentary flows \citep{Smith11a}.

Filamentary structure is clearly a key ingredient of any theory of star formation, and yet its true nature remains unexplained. One striking observation is that of \citet{Arzoumanian11} who propose that filaments have a constant width of 0.1 pc in the \herschel Gould Belt survey. We addressed this in the preceding paper in this series, \citeauthor{Smith14b}~(2014b), which investigated the projected width of filamentary column density features in turbulent clouds. However, in this work we will focus on understanding the velocity structure of the gas in and around filaments, something that is crucial for understanding how filaments may assemble the mass needed for star formation. 

\citet{Hacar13} showed that although star-forming filaments appear to be continuous features when viewed in terms of column density, they actually split up into a network of smaller sub-filaments, or `fibres' when examined in position-position-velocity space. \citet{Tafalla15} further developed this analysis and showed that these fibres were coherent velocity structures with only small velocity oscillations along their length. Together they proposed a `fray and fragment' scenario where large filaments were formed first, which then fragmented into fibres, with only the densest of these fibres collapsing to form stars. 

In this paper, we show that the formation of a network of coherent sub-filaments is a natural consequence of hydrodynamical turbulence. This filamentary network produces multiple narrow line components in C$^{18}$O emission as observed by \citet{Hacar13}. We define our terminology as follows. A filament, or filament network, is a physically related collection of sub-filaments. A sub-filament is a smaller structure, here identified using 3D positions and densities, which is part of the larger filament. The connected sub-filaments will tend to blend together into a single extended filament when viewed in column density, as was the case in the previous paper in this series. Filamentary structure is used to mean any elongated structure that could be either a filament or sub-filament.

The role of filaments in star formation has been addressed in previous analytical studies considering infinite isothermal cylinders \citep{Ostriker64,Larson85,Inutsuka92,Inutsuka97,Larson05} and in numerical studies of idealised magnetised cylinders \citep{Tilley03,Hennebelle03,Seifried15} and accreting cylinders \citep{Fischera12,Heitsch13a}. However, molecular clouds have a complex density field created by supersonic turbulence \citep[e.g.][]{MacLow04, Elmegreen04,Scalo04,Klessen14} and rarely approach such idealised scenarios. A better approach is to simulate clouds as a whole and then consider the filaments arising naturally from such turbulent structures as done in \citet{Jappsen05}, \citeauthor{Smith14b}~(2014b), \citet{Hennebelle13b}, \citet{Moeckel14}, and \citet{Gomez14}. We will use this approach to investigate the filament velocity fields in molecular clouds in this paper.

The paper is structured as follows. In Section \ref{method} we outline our numerical techniques and methods. In Section \ref{results1} we present our analysis of sub-filament velocities, and in Section \ref{results2} we investigate the time evolution and star formation in the filaments. In Section \ref{discussion} we discuss how our results might explain recent observations, and their implication for star formation. Finally in Section \ref{conclusions} we give our conclusions.

\section{Method}\label{method}
\subsection{Numerical model} \label{meth:hydro}
We perform our simulations using the moving mesh code \arepo \citep{Springel10}. This is a quasi-Lagrangian code that aims to utilise the strengths of both smoothed particle hydrodynamics (SPH) and grid-based adaptive mesh refinement (AMR) codes. The fluid is represented by a series of irregular mesh cells that are the Voronoi tessellation of a set of mesh-generating points. These mesh-generating points and the resulting mesh cells attempt to move with the flow, much as the individual particles do in an SPH simulation.  However, as the mesh is not completely Lagrangian, there is generally some residual flux of mass, momentum and energy into or out of the cells. These fluxes are computed using a Riemann solver, thereby avoiding the need to introduce artificial viscosity and allowing sharp discontinuities in the flow, such as shock fronts, to be modelled with a thickness of only 1--2 mesh cells. The resulting mesh is adaptable and can be refined to give improved resolution in regions of interest. This allows the study of problems with an extreme dynamic range, that are discontinuous, and that involve fluid instabilities, all while imparting no preferred geometry on the problem.

We model the chemical evolution of the gas using the NL97 network described in \citeauthor{Glover12a}~(2012a), which combines the hydrogen chemistry of \citet{Glover07a,Glover07b} with the simplified treatment of CO formation and destruction introduced in \citet{Nelson97}. We assume that the strength and spectral shape of the ultraviolet portion of the interstellar radiation field (ISRF) are the same as the values for the solar neighbourhood derived by \citet{Draine78}; note that this corresponds to a field strength of 1.7~ \citet{Habing68} units of $1.6 \E^{-3}$ erg cm$^{-2}$ s$^{-1}$. We also include the effects of cosmic rays and adopt a rate $\zeta_{\rm H} = 3 \times 10^{-17} \: {\rm s^{-1}}$ for atomic hydrogen, and a rate twice the size of this for molecular hydrogen. This model was first implemented in \arepo in  \citeauthor{Smith14a}~(2014a).

To treat the attenuation of the ISRF due to H$_{2}$ self-shielding, CO self-shielding, the shielding of CO by H$_{2}$, and by dust absorption, we use the \treecol algorithm developed by \citet{Clark12b}. This algorithm computes a $4\pi$ steradian map of the dust extinction and H$_{2}$ and CO column densities surrounding each \arepo cell, using information from the same oct-tree structure that \arepo uses to evaluate gravitational interactions between cells. The resulting column density map is discretised onto $N_{\rm pix}$ equal-area pixels using the {\sc healpix} pixelation algorithm \citep{healpix}. In the simulations presented here, we set $N_{\rm pix} = 48$. To convert from H$_{2}$ and CO column densities into the corresponding shielding factors, we use shielding functions taken from \citet{Draine96} and \citet{Lee96}, respectively. We assume that the radiation field is uniform and enters through the sides of the box. The inclusion of time-dependent chemistry allows us to calculate the radiative heating and cooling of the gas self-consistently within the simulation. It is important to model this accurately, as we expect filament formation to be significantly easier when the effective equation of state of the gas is sub-isothermal \citep{Larson85, Peters12b} and so the use of a simple isothermal or polytropic equation of state may lead to unrepresentative results.

\subsection{Simulations}\label{meth:sim}

In the previous paper in this series (\citeauthor{Smith14b}~2014b), we considered a set of simulations with different turbulent properties. However, in this paper we chose to focus on the two simulations containing a natural mix of solenoidal and compressive turbulence \citep[e.g.][]{Schmidt09,Federrath10b}. In both of these simulations, we model $10^4$\msun gas clouds that have solar metallicity and that are initially composed of fully atomic gas. The initial condition is that of a uniform sphere with a hydrogen nucleus number density of $n \sim 100 \: {\rm cm^{-3}}$ and a radius of 7~pc. This is embedded in a larger 65 pc periodic box containing a tenuous warm medium with a temperature of several thousand Kelvin. The size of this larger box is such that the dense cloud at its centre never encounters the box edges. This setup is analogous to an isolated molecular cloud in the ISM. The turbulence has a $P(k)\propto k^{-4}$ power spectrum (specifically, the same spectrum as in runs S1 and S2 in the nomenclature of Smith et~al., 2014b). The magnitude of the root mean square turbulent velocity is normalised such that the clouds have an equal amount of kinetic and gravitational potential energy at the start of the simulations. Within each simulation we focus on the filaments within two different regions, and then follow the time evolution of one of these regions over five different snapshots. This gives us twelve 3.25 pc$^3$ regions to which we can apply our filament finding algorithm.

In order to ensure that the filaments are well resolved on all scales we require the Jeans length to be resolved by a minimum of 16 cells at all densities. This ensures that there is no artificial fragmentation in the gas, and that the filament velocities are well resolved at their centres.  Where necessary, the grid is refined until the condition is satisfied. We calculate this Jeans length assuming a fixed temperature of 10 K for the gas. Very little of the dense gas is colder than this, and most is at least a few K warmer, and so this gives us a conservative estimate for the size of the Jeans length. A full resolution study of the simulations used here was carried out in \citeauthor{Smith14b}~(2014b), which also shows the dependence of the resolution on the gas density. For example, at densities of $10^4$ \cmc~the cell radius is $\sim 5\E^{-3}$ pc, and at our highest densities of $10^7$ \cmc~the cell radius is $2\E^{-4}$ pc.
%%% RSK: changed radius into diameter: please check!

Star formation is modelled in the simulation using sink particles \citep{Bate95}. These were first introduced into \arepo by \citet{Greif11} and we use a slightly modified version of this routine here. Above number densities of $10^7$\cmc, we check whether the densest cell in the simulation and its neighbours satisfy the following three conditions: 1) the cells are gravitationally bound, 2) they are collapsing and 3) the divergence of the accelerations is less than zero, so the particles will not re-expand (see also \citealt{Federrath10a}). If all these conditions are satisfied, the cell and its neighbours are replaced with a sink particle, which interacts with the gas cells purely through gravitational forces. Additional material can be accreted by the sink particles if it is within an accretion radius of $r_{\rm acc}= 0.01$ pc (which corresponds to 25 resolution elements at our chosen sink creation density) and is bound to the sink.  We use relatively large sink particles in this study in order to focus on the geometry of the filaments without interference from very dense collapsing cores which would distort the average of the filament profile. Consequently the sinks should not be thought of as `stars', but as collapsing cores. 

\subsection{Filamentary structure identification}\label{meth:fil}

\begin{figure}
\begin{center}
\includegraphics[width=3 in]{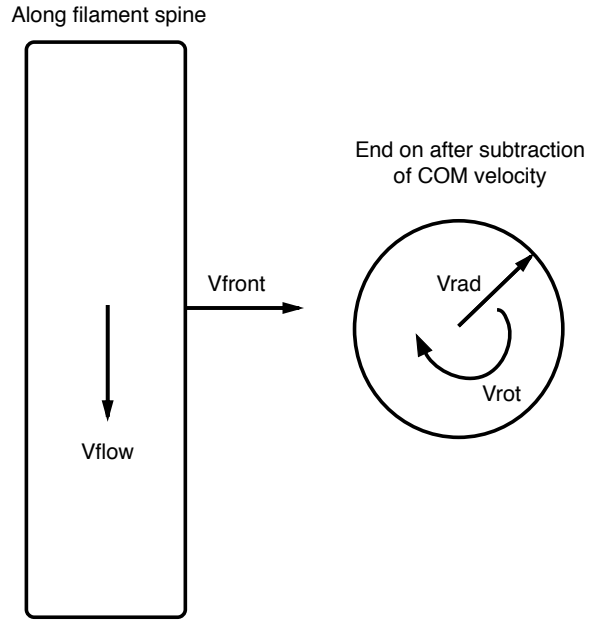}
\caption{A simple cartoon showing how we decompose the sub-filament velocities. The flow velocity is the velocity along the spine of the sub-filament, and the front velocity is the vector sum of the velocity components perpendicular to the spine. The radial and rotational velocities are the \textit{residual velocities} perpendicular to, and around the spine after the mean velocity has been subtracted.}
\label{cartoon}
\end{center}
\end{figure}

The filamentary structures we study are identified using the \disperse (DIScrete PERsistent Structures Extractor) algorithm \citep{Sousbie11}. This algorithm constructs a Morse-Smale complex from an input density distribution and identifies the critical points where the density gradient is zero. These can then be connected to delineate structures in the data. Filamentary structures are found by connecting the points such that maxima are connected to saddle-points along Morse field lines. To avoid artificial structures being identified, we extract only the the structures which have a persistence ratio with a probability of 5 sigma or more when compared to Poisson noise. We apply \disperse to uniform 3D density grids of regions where filaments were found in \citeauthor{Smith14b}~(2014b) and extract the grid points where filamentary structures with a density greater than some threshold occur ($2.5\E^{3}$ \cmc~ for the bulk of our cases). We use this `skeleton' network to get a series of vectors (hereafter called segments) describing the orientation of each filamentary structure. The structures do not have to be linear, but instead can bend and twist. For the co-ordinates of the segments (referred to as spine points), we use the position of the densest \arepo cell within a resolution element of the \disperse input grid for each skeleton point.

The filament skeleton map obtained above does not contain any information about the properties of the filamentary structures, so we have to calculate these from the \arepo simulation. For each segment found by \disperse, we find the mass of the segment by measuring the mass contained within a cylinder of radius $r = 0.035$~pc oriented along the vector. This value was chosen for the radius because this was found to be the typical flattening radius of a density sub-filament in our previous paper. To find the flattening radius in \citeauthor{Smith14b}~(2014b) we fit a Plummer-like function which describes the radial variation in the number density $n(r)$, 
\begin{equation}\label{Plum-dens}
n(r)=\frac{n_{\rm c}}{[1+(r/R_{\rm flat})^2]^{p/2}} + B_{\rm 3D} [{\rm cm}^{-3}]\textcolor{red}{~~,}
\end{equation}
where $n_{\rm c}$ is the central density, $R_{\rm flat}$ is a radius within which the profile is flat, $p$ sets the slope of the power law fall-off beyond this radius, and $B_{\rm 3D}$ represents the background density. This is a common parameterisation of filamentary density profiles and is widely used in the literature \citep[e.g][]{Nutter08,Arzoumanian11,Kirk15}. The typical mean values for the sub-filaments are $n_c\sim10^6$ \cmc, $R_{\rm flat}\sim0.035$ pc, and $p\sim 1.9$. While we concentrate  on dense star-forming filaments in \citeauthor{Smith14b}~(2014b), in the current study we also include more diffuse filamentary regions.

To find the total mass of the filamentary structure we add together the mass in each segment within our chosen radius. Similarly, we add the lengths of the individual segments to get the total filament or sub-filament length. We assign a net velocity to each spine point by taking the mass-weighted average velocity of all the \arepo cells within this cylinder. This means that we are only including the gas at the very centre of the sub-filaments where the density profile is flat. A different choice of radius, would of course, lead to different sub-filament masses and accretion rates. However, we chose to use the mean value of $R_{\rm flat}$ from \citeauthor{Smith14b}~(2014b) since it is a physically motivated choice that is small enough that we can be sure only a single sub-filament will be included in the mass estimate, and not any nearby neighbour. It should be stressed that the total mass of an equivalent filament seen in observations will be higher than this value, since it will include the total of all the sub-filaments, plus a contribution from the more diffuse gas seen in column density out to a larger radius. 

In order to investigate the velocity structure of the sub-filaments, we decompose the velocities along their skeletons into the four components shown in \fig \ref{cartoon}. The flow velocity $v_{\rm flow}$ is the mass-weighted velocity along the sub-filament segment, and the front velocity $v_{\rm front}$ is the mass-weighted velocity perpendicular to the sub-filament segment. These centre of mass velocity components are then subtracted from the velocity in each \arepo cell to find the \textit{residual} velocities around the sub-filament spine. The radial velocity $v_{\rm rad}$ is the mass-weighted velocity measured radially outward from the sub-filament centre, as shown in \fig \ref{cartoon}. When calculating $v_{\rm rad}$ we include \arepo cells out to a radius of 0.15~pc so that gas on the power-law part of the density distribution is also included in order to see if mass is flowing towards or away from the centre. Finally, the rotational velocity $v_{\rm rot}$ is the mass-weighted velocity component perpendicular to $v_{\rm rad}$, which describes whether the sub-filament is spinning about its long axis.

\section{Sub-Filament Velocities}\label{results1}

\subsection{Filament Maps}

\begin{figure*}
\begin{center}
\begin{tabular}{c c c}
\begin{overpic}[scale=0.38]{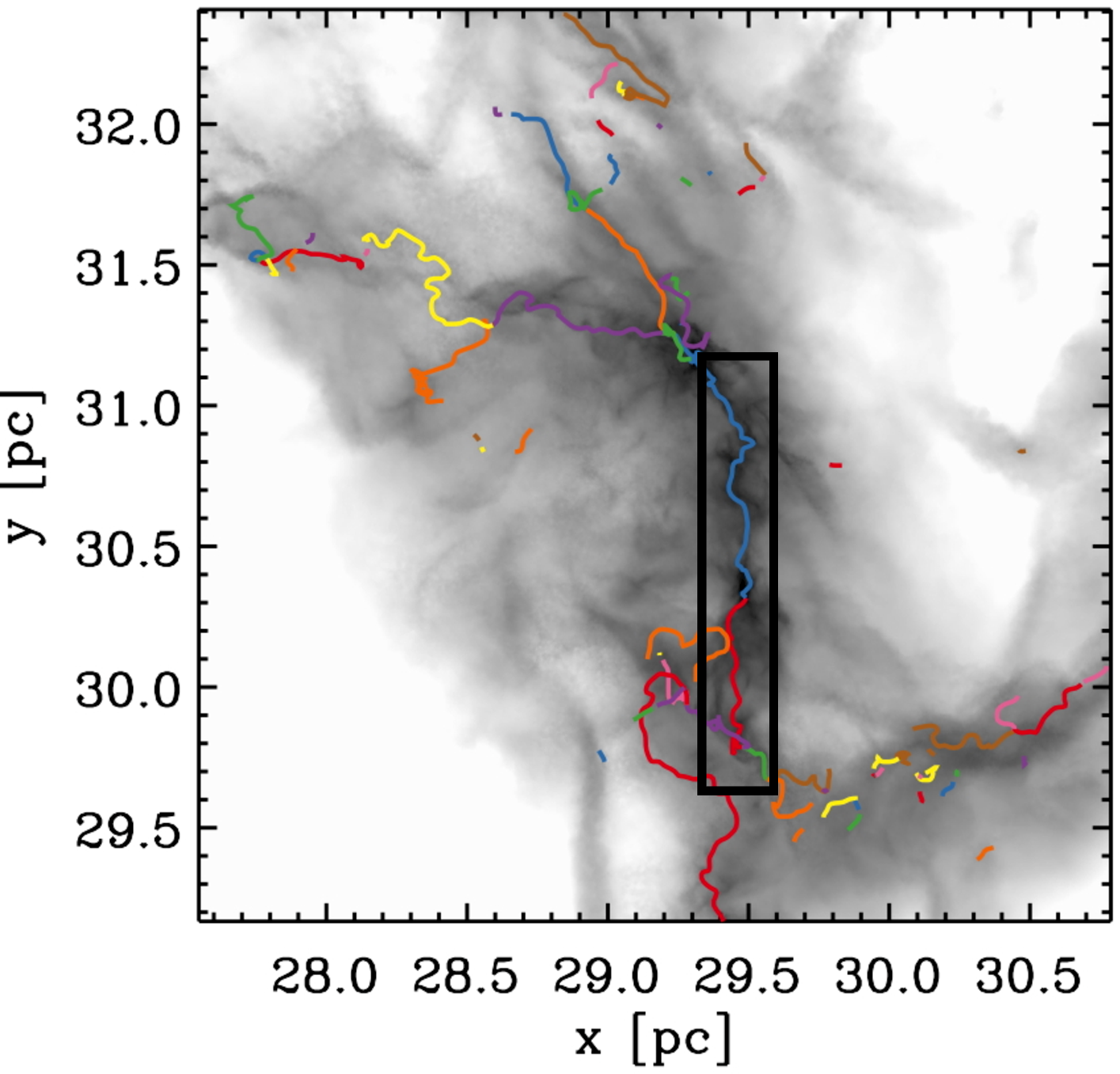}
\put (75,90) {\makebox(0,0){{\bf S1F1T303}}}
\end{overpic}
\begin{overpic}[scale=0.38]{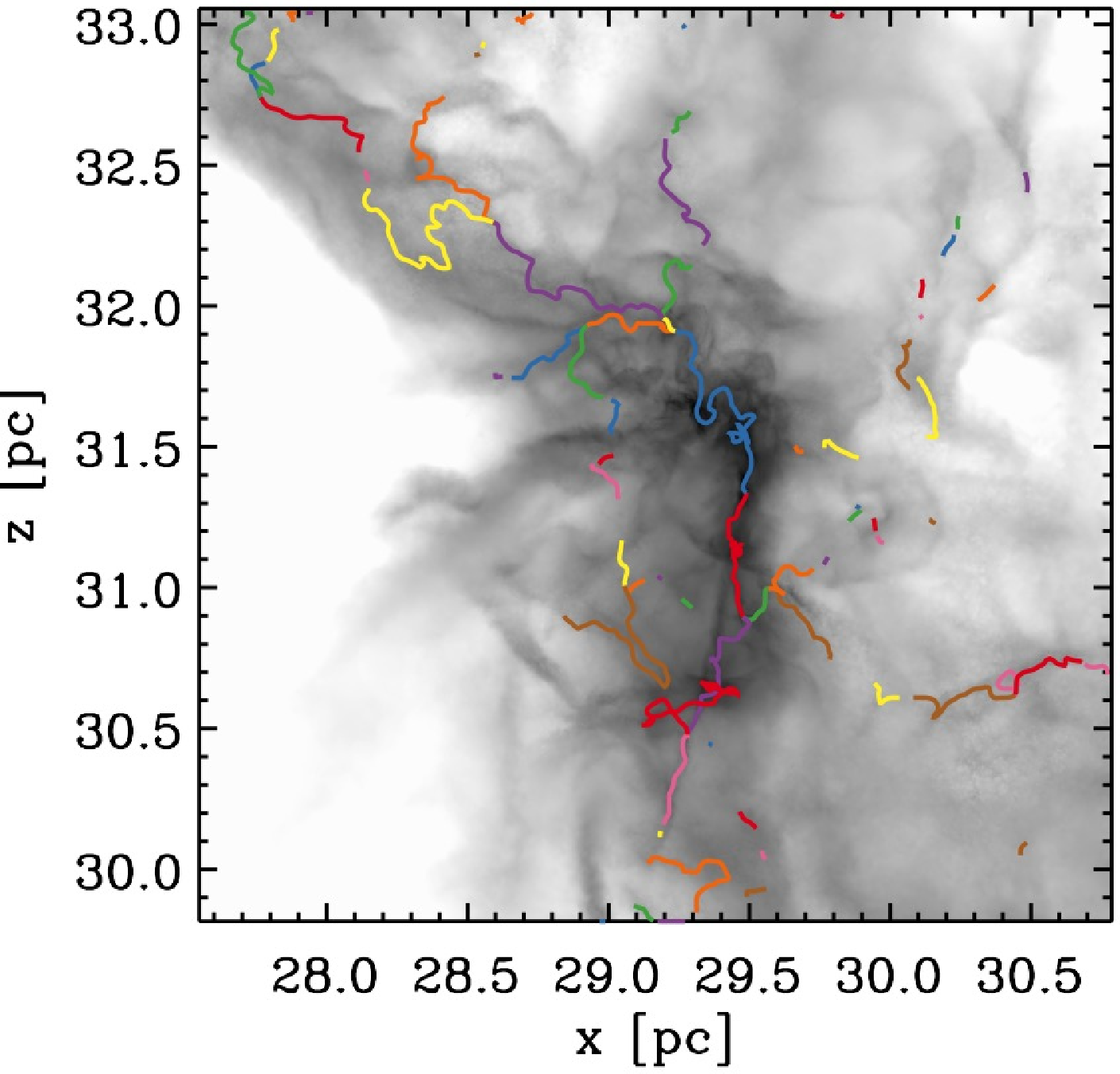}
\put (75,90) {\makebox(0,0){{\bf S1F1T303}}}
\end{overpic}
\begin{overpic}[scale=0.38]{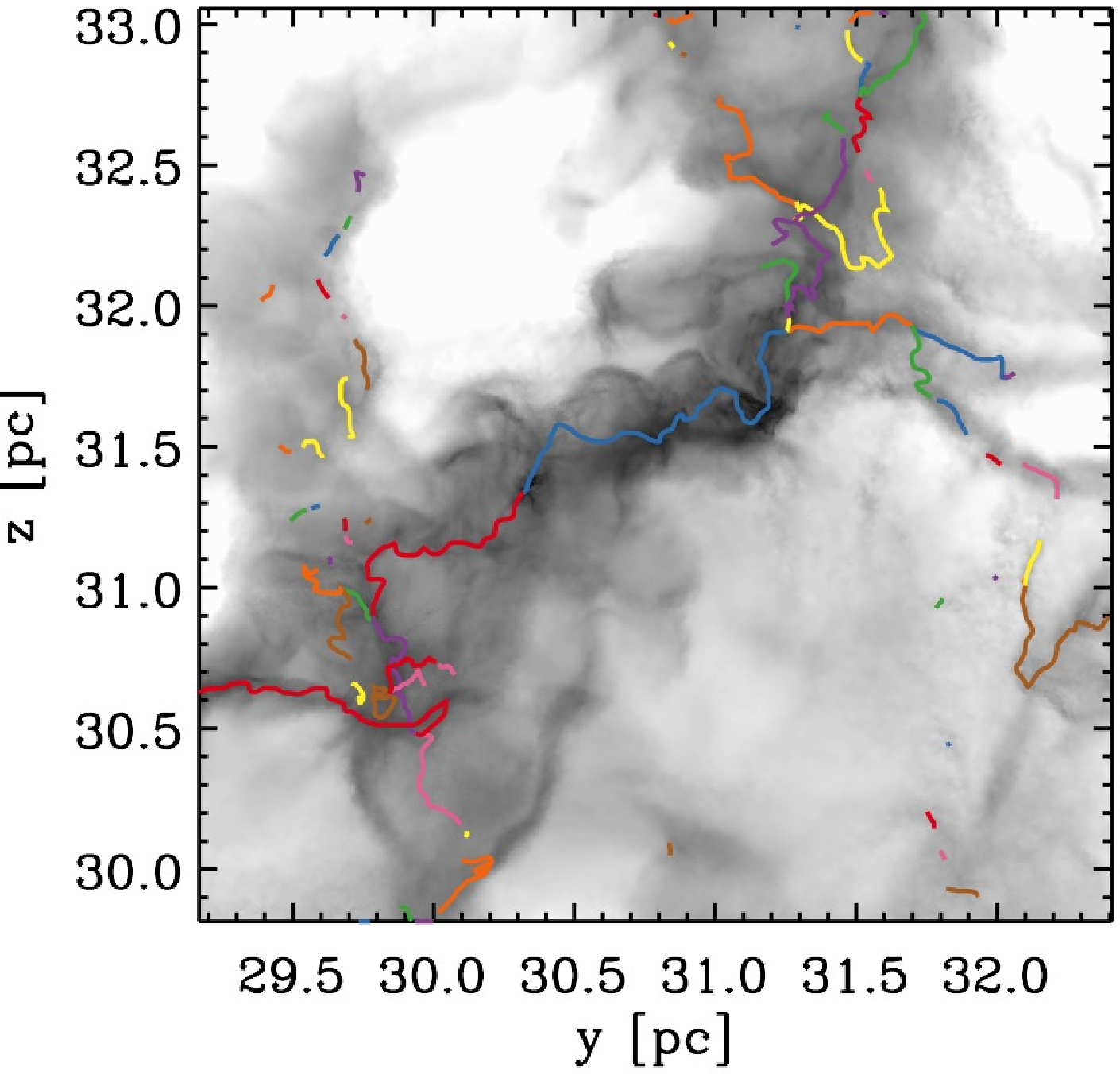}
\put (75,90) {\makebox(0,0){{\bf S1F1T303}}}
\end{overpic}\\

\begin{overpic}[scale=0.38]{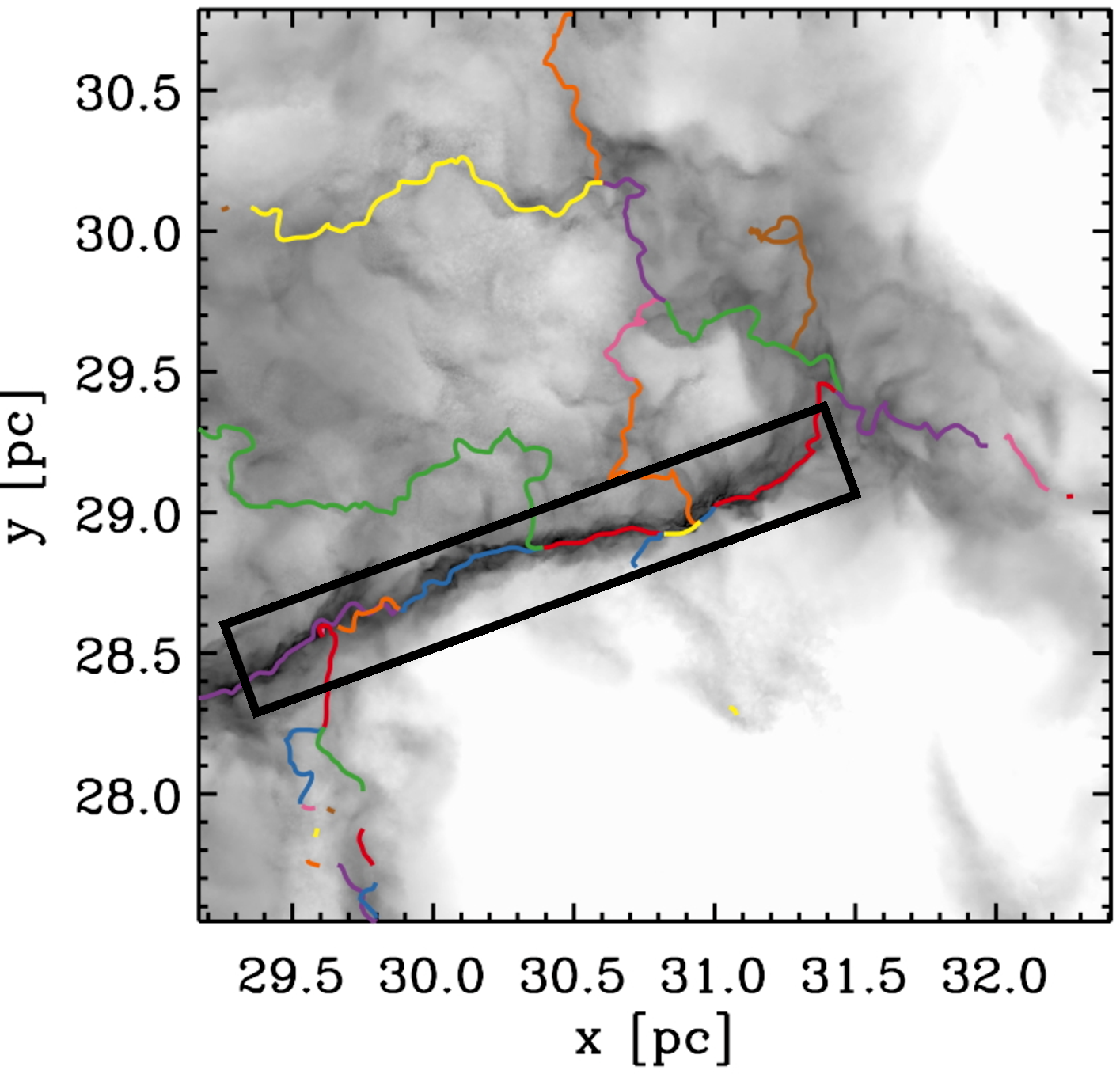}
\put (75,90) {\makebox(0,0){{\bf S2F1T390}}}
\end{overpic}
\begin{overpic}[scale=0.38]{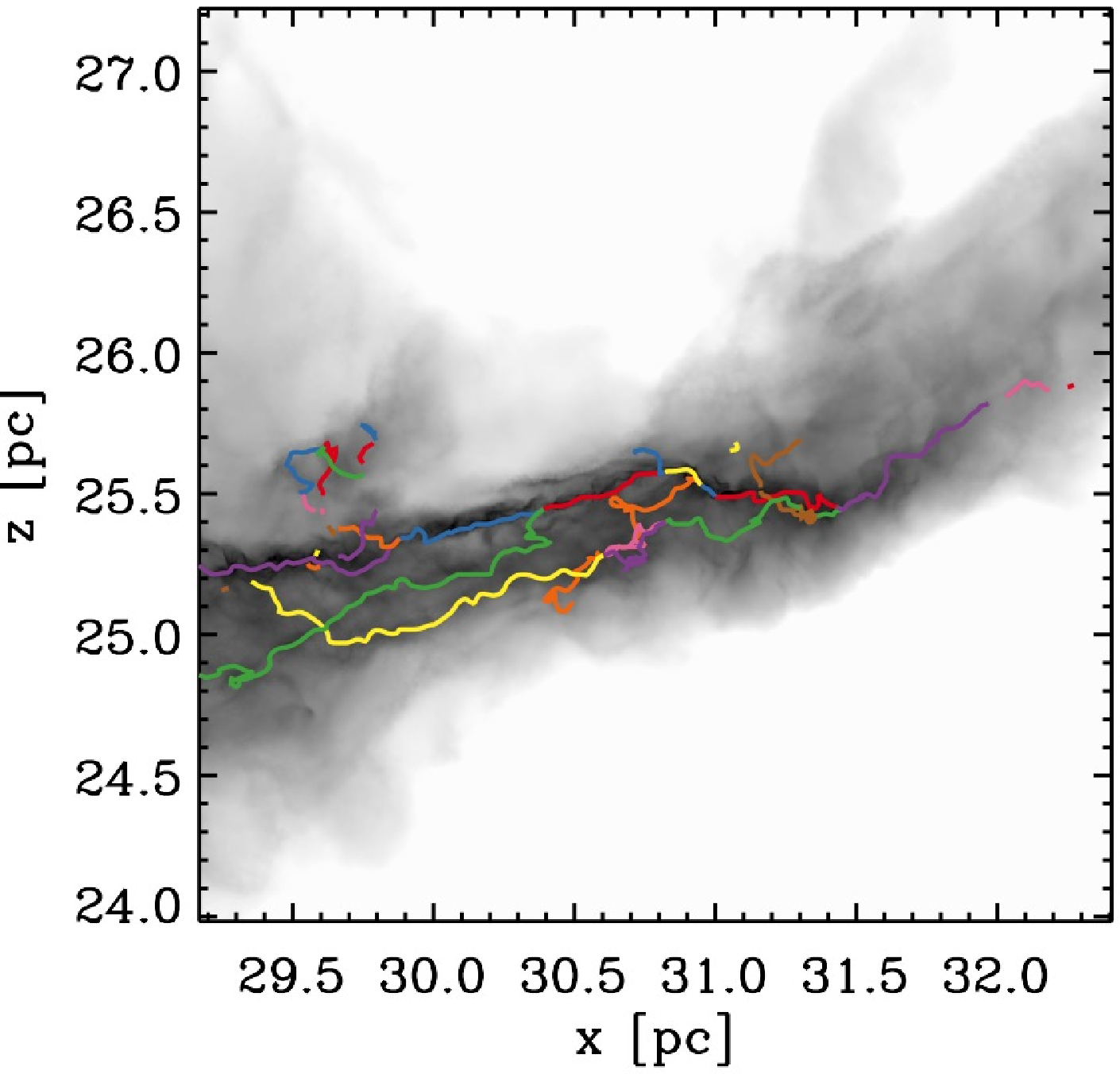}
\put (75,90) {\makebox(0,0){{\bf S2F1T390}}}
\end{overpic}
\begin{overpic}[scale=0.38]{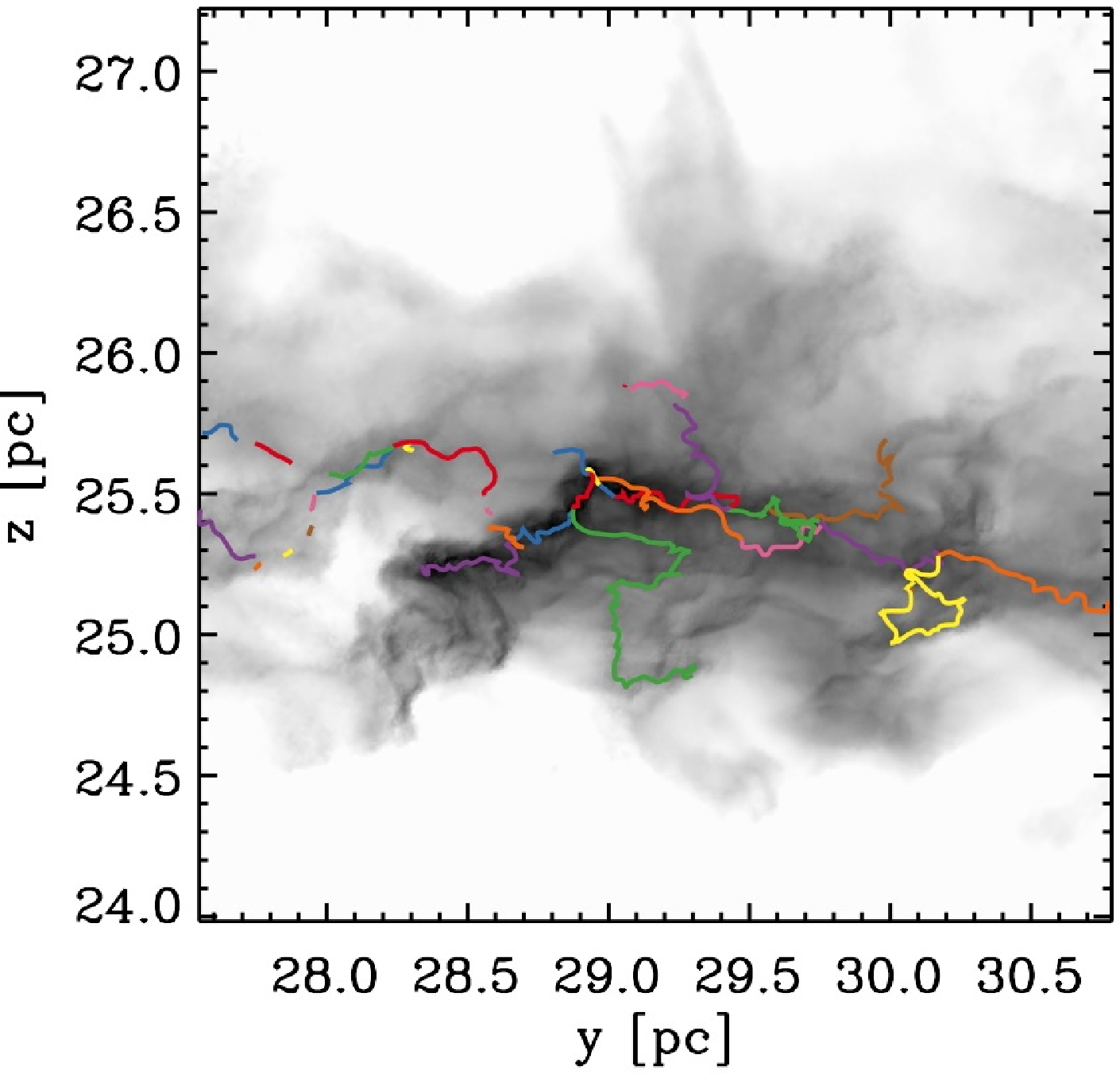}
\put (75,90) {\makebox(0,0){{\bf S2F1T390}}}
\end{overpic}\\

\begin{overpic}[scale=0.38]{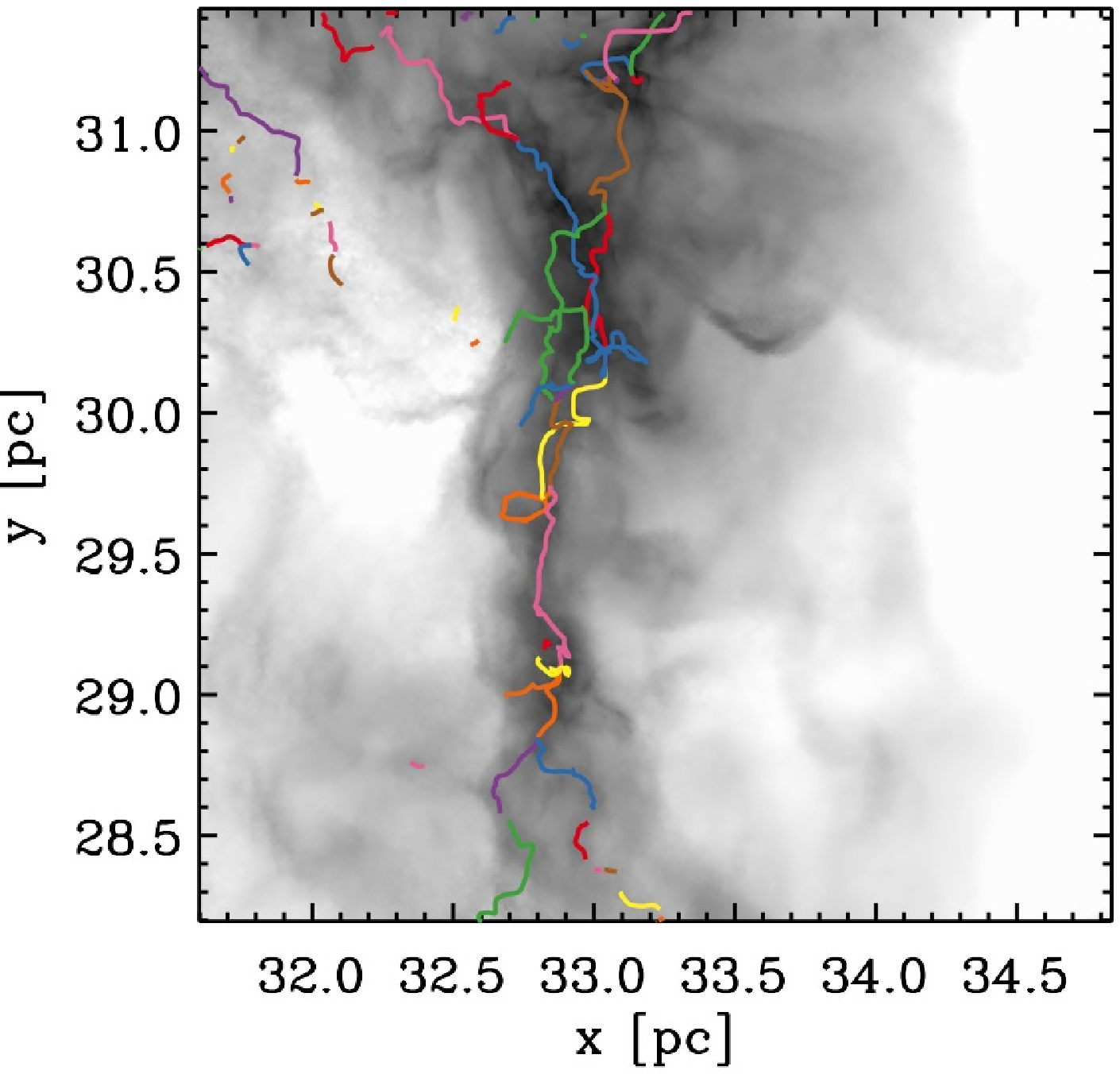}
\put (75,90) {\makebox(0,0){{\bf S1F2T303}}}
\end{overpic}
\begin{overpic}[scale=0.38]{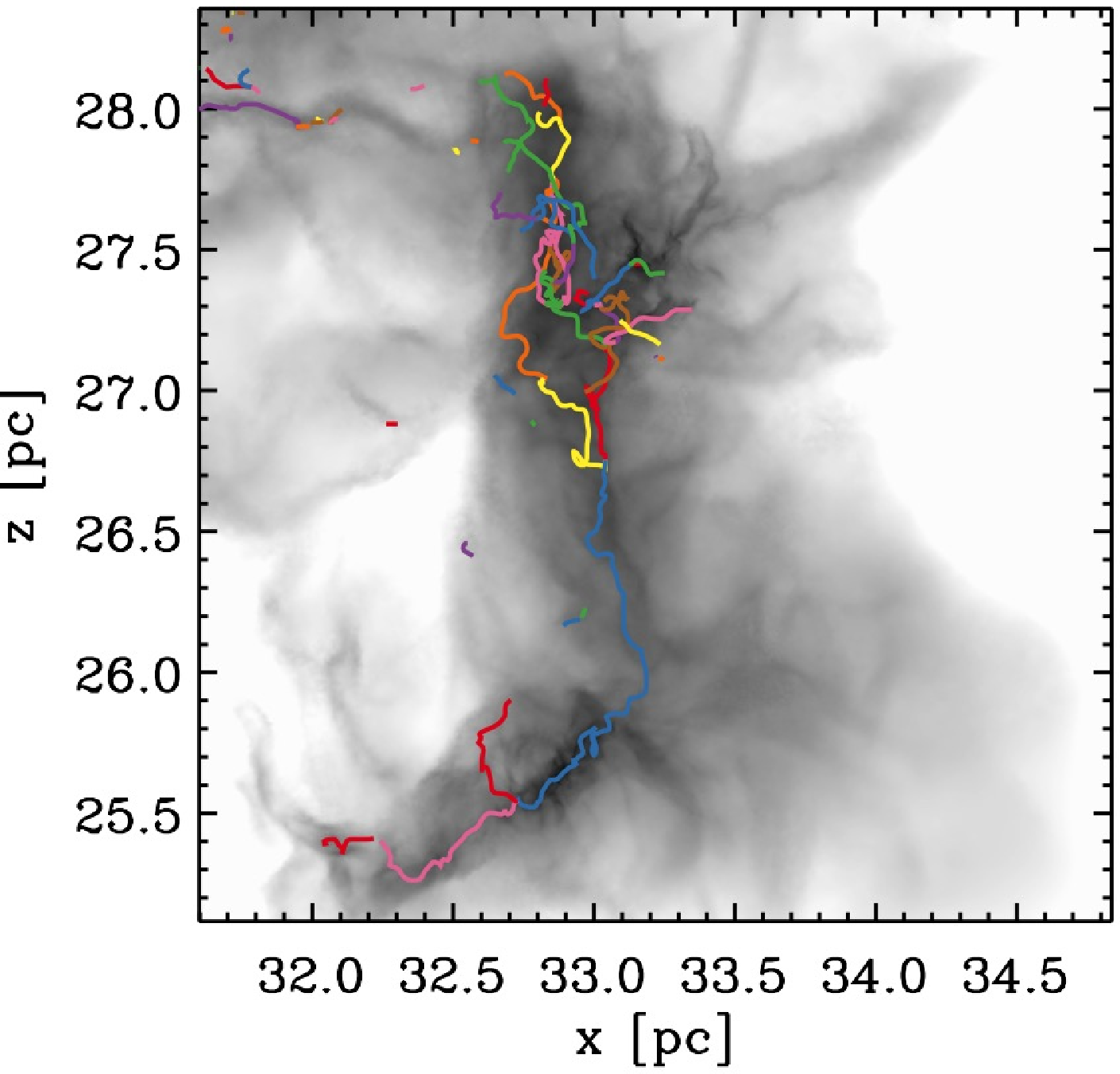}
\put (75,90) {\makebox(0,0){{\bf S1F2T303}}}
\end{overpic}
\begin{overpic}[scale=0.38]{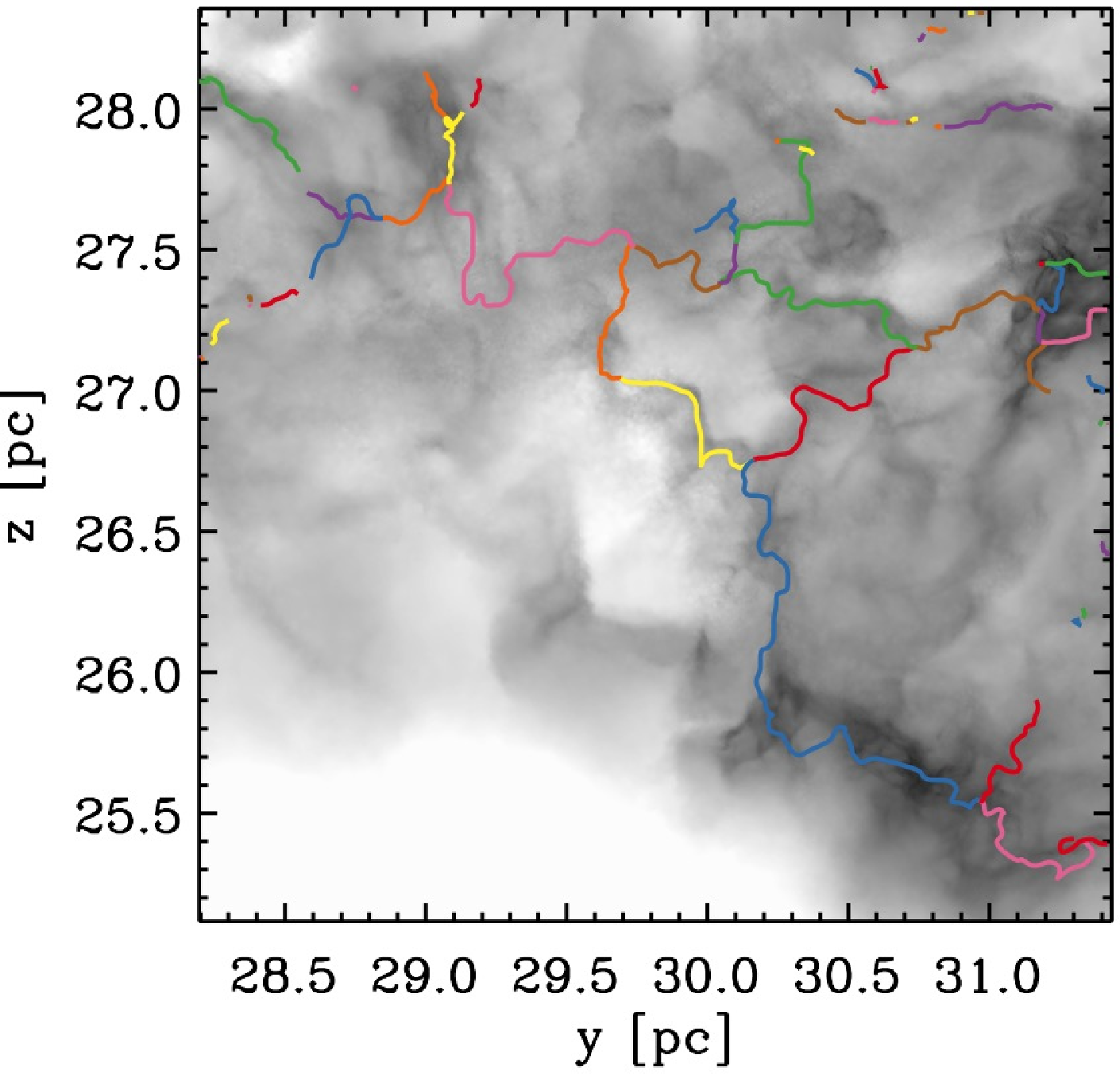}
\put (75,90) {\makebox(0,0){{\bf S1F2T303}}}
\end{overpic}\\

\begin{overpic}[scale=0.38]{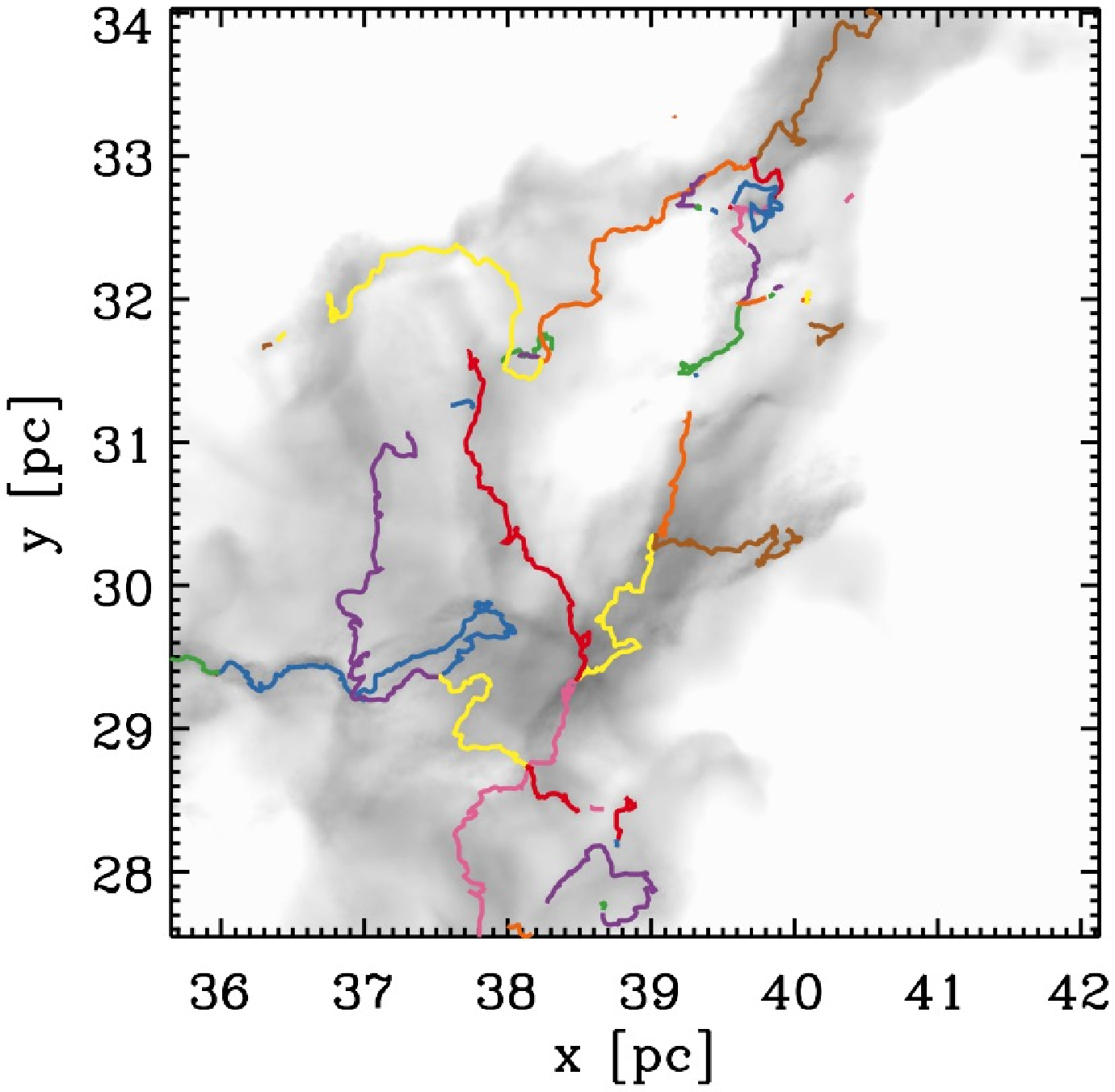}
\put (75,90) {\makebox(0,0){{\bf S2F2T216}}}
\end{overpic}
\begin{overpic}[scale=0.38]{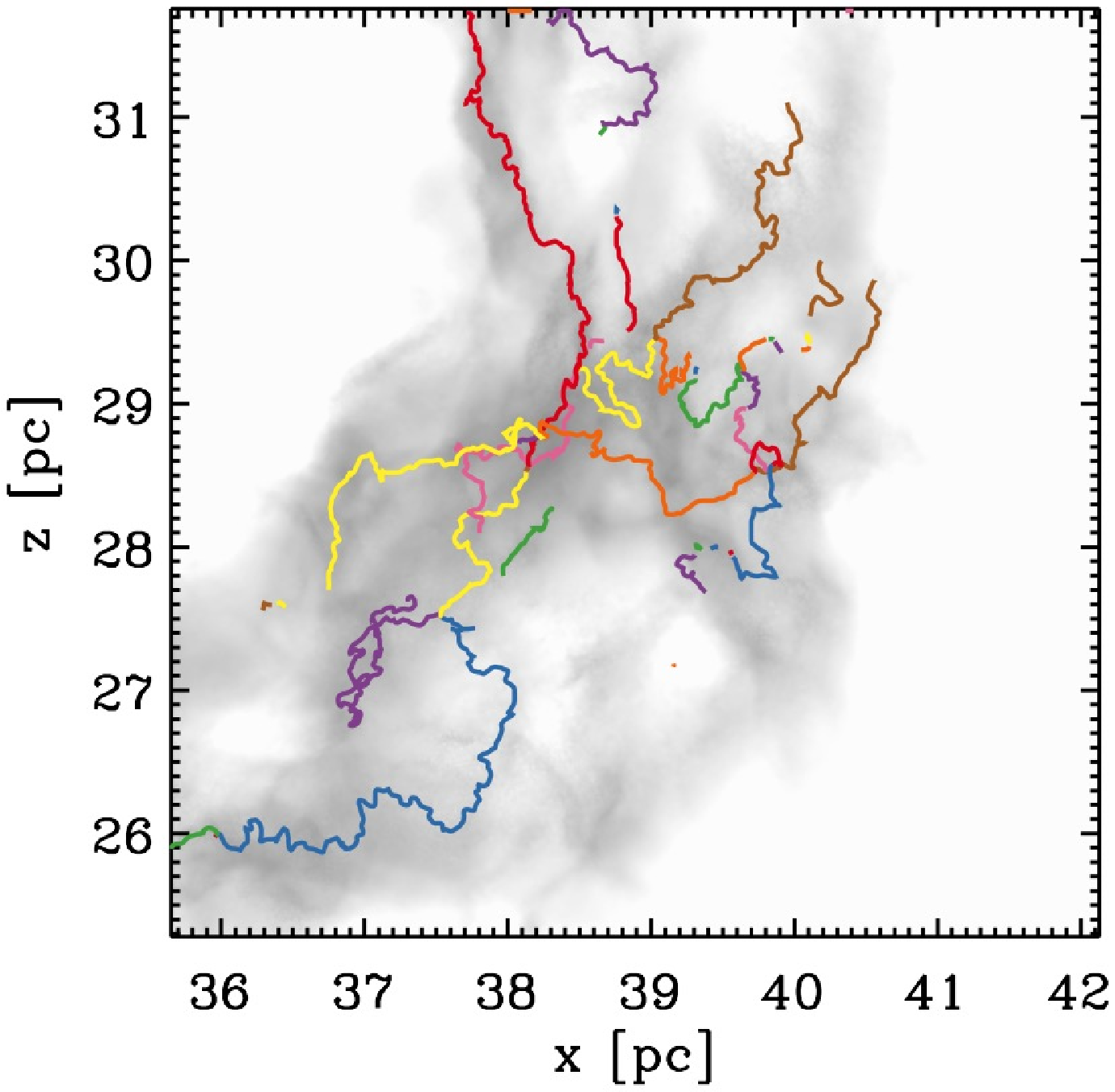}
\put (75,90) {\makebox(0,0){{\bf S2F2T216}}}
\end{overpic}
\begin{overpic}[scale=0.38]{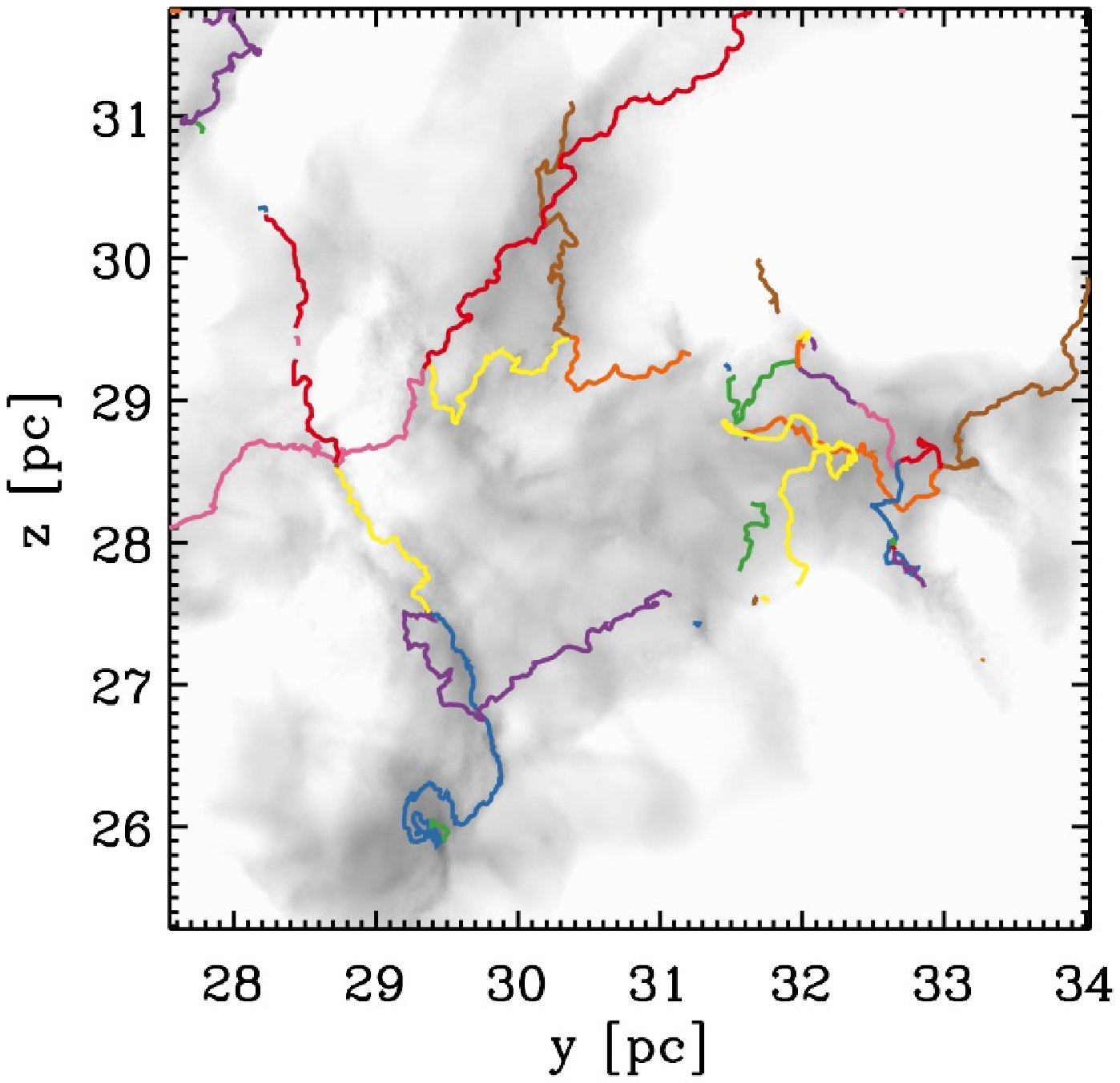}
\put (75,90) {\makebox(0,0){{\bf S2F2T216}}}
\end{overpic}\\
\end{tabular}
\caption{A column density projection of the four different regions within our simulations that we use for our analysis. The column density is shown in greyscale and the filament skeleton obtained from \disperse is plotted on top. Different sub-filaments are shown in different colours so that they can be distinguished. The black boxes show the sub-filaments that are analysed in Figure \ref{flow_spine}.}
\label{columns}
\end{center}
\end{figure*}

Figure \ref{columns} shows the column density of four filaments with the skeleton obtained from \disperse plotted on top.The filaments are named using the convention where S denotes the number of the simulation, F denotes the filament within the simulation, and T denotes the time in units of $10^4$ yr since the beginning of the simulation. The filament skeletons were cut off below a density of $2.5\E^{3}$ \cmc~ for S1F1, S2F1 and S1F2, but in the case of S2F2 they were cut off at $2.5\E^{2}$ \cmc. This region was chosen to be larger and more diffuse to contrast with the other three cases where some of the sub-filaments eventually form stars. The sub-filament spines generally line up well with the column density projection, but occasionally a feature is seen in column density that is not identified as a sub-filament spine because its density is not high enough to meet our selection criteria. For example, a long tenuous density feature may have a high column density but not have an intrinsic density above our threshold.

When viewed in another projection the distributions look quite different, showing the 3D nature of the filament network. If one were to view the column density in the xy plane seen in the left panels, then the z direction would be the observer's line of sight. In the panels we see that there is the potential for multiple sub-filaments to contribute to the emission from a single observed pointing. These overlapping sub-filaments are very reminiscent of the observations of \citet{Hacar13}, as we shall discuss further in Section \ref{discussion}

%While \fig \ref{columns} only shows the column density projections, in the online material we include movies which show density slices through S1F1 and S2F1 with velocity vectors. These reveal that what appears to be a single filament in column density is actually a projection of a 3D network of sub-filaments, as we discussed in Paper I. Hereafter, we will refer to the filaments in our simulated boxes as sub-filaments, as they appear to be part of a single filament when viewed in column density. %The movies also reveal that sub-filaments are preferentially formed at the divergence points of the velocity field, in particular where there is a converging flow, or shearing flows that will stretch the gas into elongated structures.
 %In the Appendix we show the same plot, but viewed in the $x$-$z$ plane rather than the $x$-$y$ plane. 

\subsection{Front and Flow Velocities}

\begin{figure*}
\begin{center}
\begin{tabular}{c c}
\begin{overpic}[scale=0.4]{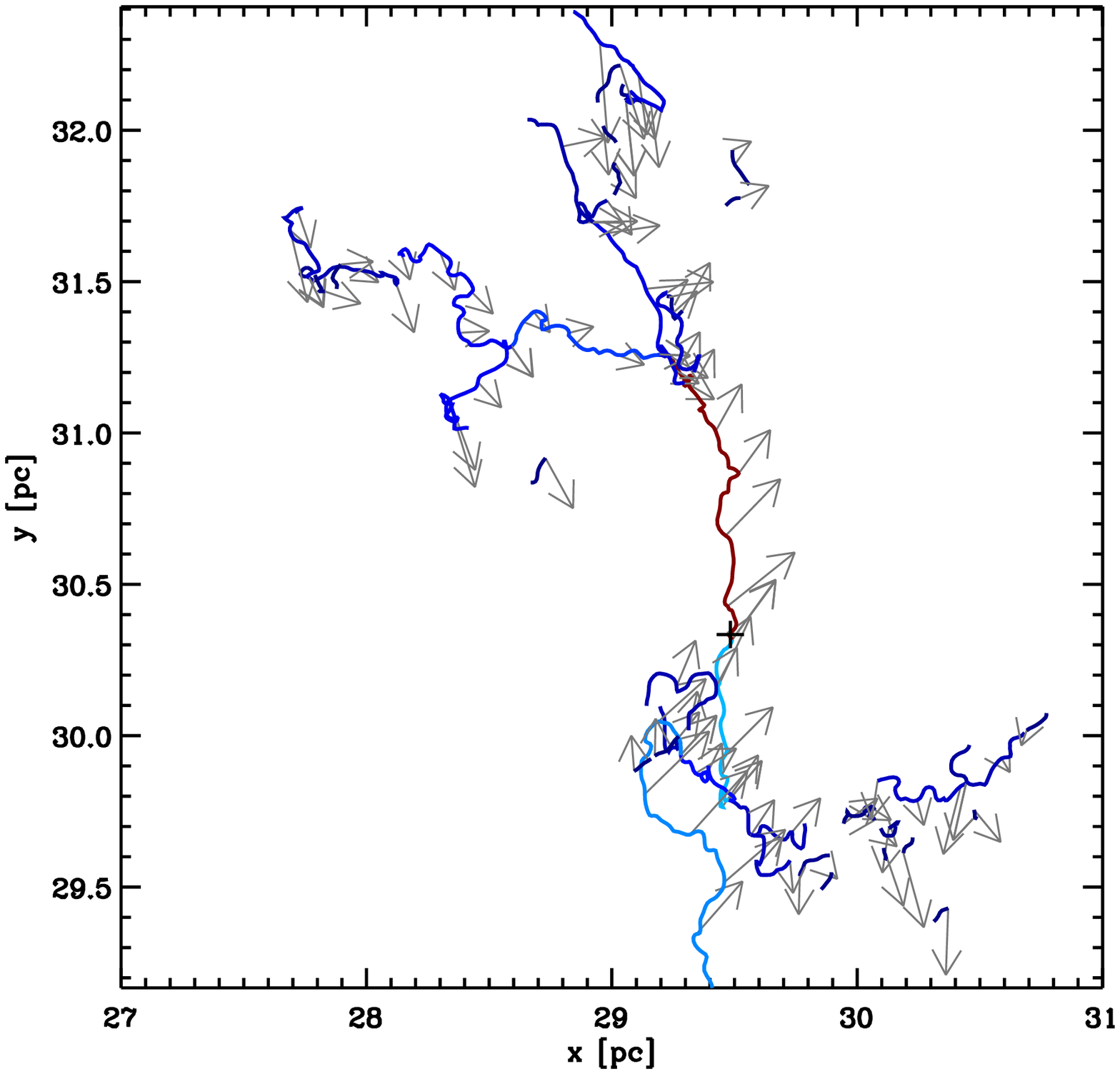}
\put (75,90) {\makebox(0,0){{\bf S1F1T303}}}
\end{overpic}
\begin{overpic}[scale=0.4]{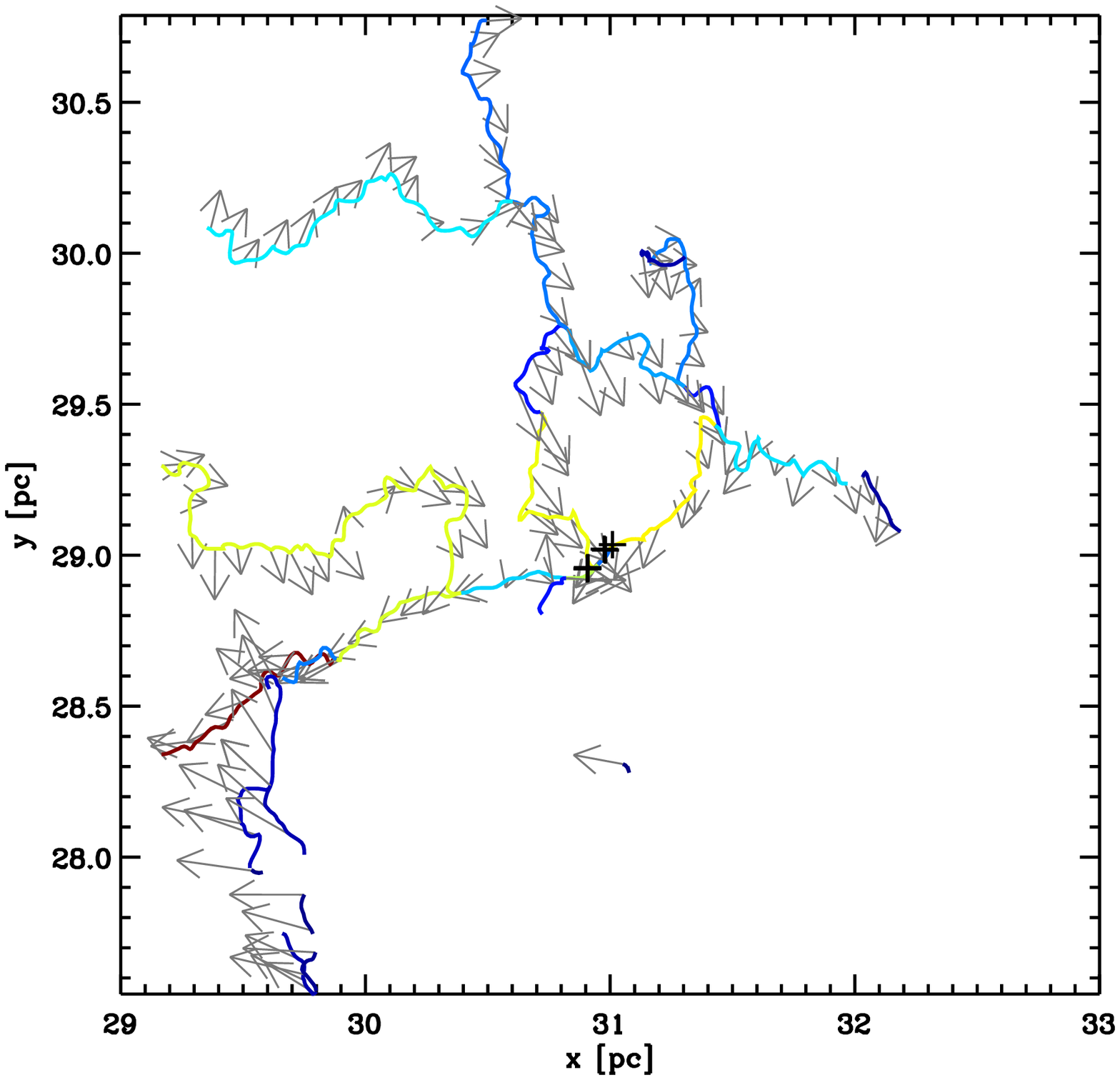}
\put (75,90) {\makebox(0,0){{\bf S2F1T390}}}
\end{overpic}\\
\end{tabular}
\caption{An illustrative example of the velocities at points along the sub-filament spine for two of our filament networks. The colours denote the mass in each sub-filament, with dark blue being the least massive and red the most. The grey arrows show the direction of the mass-weighted velocities at points along the sub-filament spines. The black crosses show where sink particles are being formed.}
\label{skel_xy}
\end{center}
\end{figure*}

In Figure \ref {skel_xy} we show two of the filament skeleton networks, but this time colour-code the skeletons so that the most massive sub-filaments are red and the least massive are dark blue. Arrows show the mean velocity direction and relative strength at every tenth spine point. Black crosses show where sink particles, representing star-forming cores, are formed. Figure \ref{skel_xy} clearly shows that the filament networks are dynamical features with significant bulk motions.

A better understanding of how the density sub-filaments correspond to the turbulent field can be seen in the online material \footnote{ movies are available at http://www.jb.man.ac.uk/$\sim$rjs/S1\_T130\_g400.mov and http://www.jb.man.ac.uk/$\sim$rjs/S2\_T170\_g400.mov}. This contains movies showing the velocity and density field in slices moving through the simulated filament networks for S1F1 and S2F1. Typically sub-filaments are formed at the convergence points of the velocity field where mass is swept up by the flow and the density is increased by shocks. This is the formation mechanism of the sub-filaments seen in our simulations. In addition to the motions perpendicular to the sub-filaments (\vfront), there are also motions parallel to the sub-filaments (\vflow) which can be seen particularly clearly in the case of S2F1.

\begin{figure}
\begin{center}
\includegraphics[width=3in]{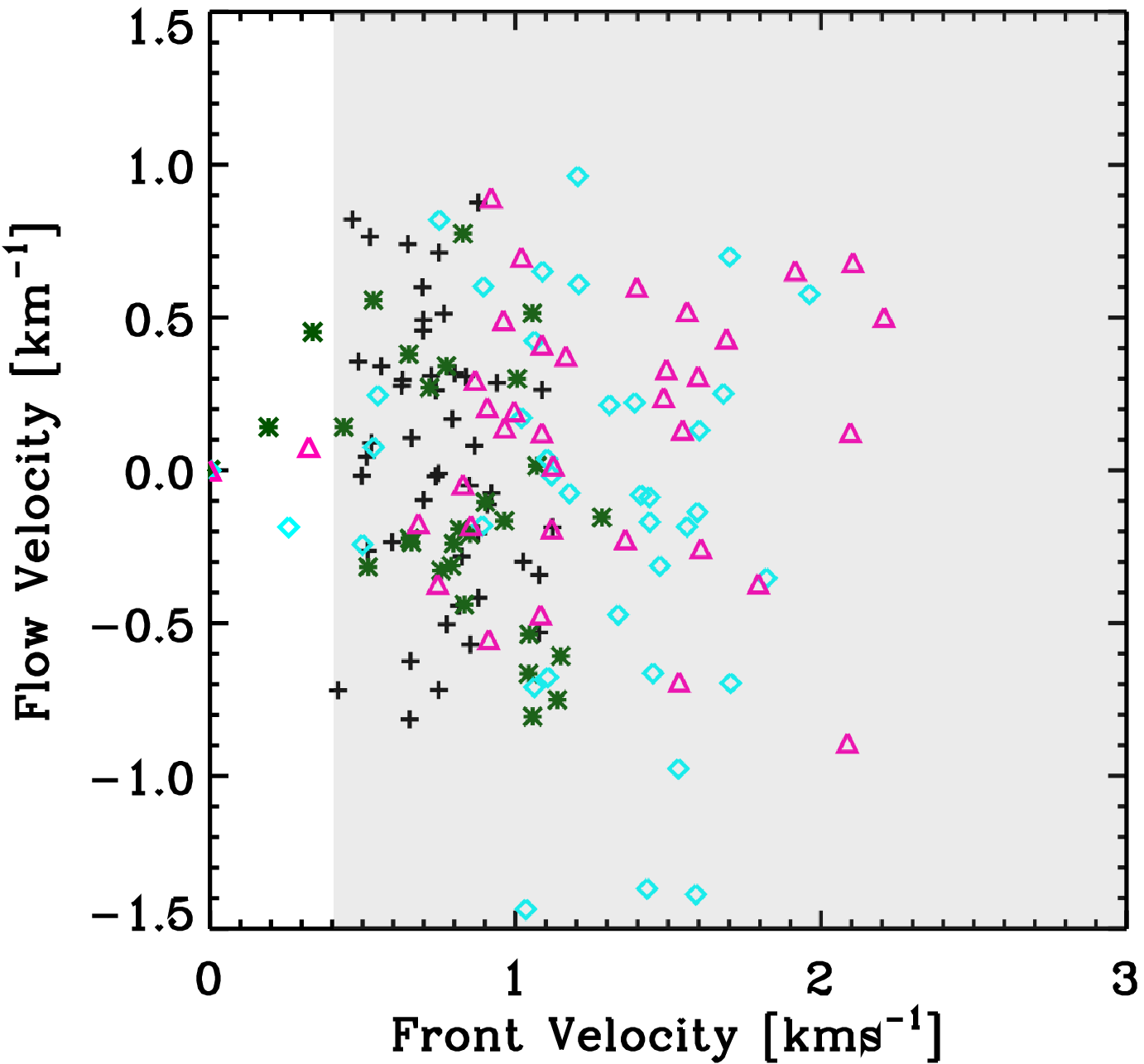}
\caption{The mean front vs flow velocity for the sub-filaments. Black crosses show sub-filaments from S1F1, green asterisks show sub-filaments from S2F1, cyan diamonds show sub-filaments from S1F2, and pink triangles sub-filaments from S2F2. The sub-filaments in the grey shaded box are travelling faster than the sound speed, which means that there will be shock fronts where the sub-filaments collide with slower moving gas.}
\label{sum_front_flow}
\end{center}
\end{figure}

In Figure \ref{sum_front_flow} we plot the mean front and flow velocities for each of the sub-filaments shown in Figure \ref{columns}. The sub-filaments are colour coded to show which simulated filament network they are part of. Table \ref{summary_table} shows the mean velocities and mass of the sub-filaments in each filament network. The ratio of the magnitudes of the front velocity to the flow velocity is consistently high. By definition the front velocity must be positive, since it is the magnitude of the two velocity vectors perpendicular to the filament long axis, and as such we would expect it to be slightly higher, by a factor of $\sqrt 2$, than the flow velocity in the case of random sampling. Instead the ratios are much larger than this value, about a factor of six, which leads to the conclusion that the sub-filaments are velocity fronts moving through the cloud. The sound speed in the simulated molecular cloud is typically less that 0.4 \kms (about 0.2 \kms~to 0.3\kms~in the dense gas). Consequently where the sub-filaments move through a static or slow-moving gas background they will be moving shock fronts.

%start here

\begin{table*}
	\centering
		\begin{tabular}{l c c c c c c c c c}
   	         \hline
	         \hline
		  Simulation & $N_{\rm fil}$ & M$_{{\rm tot}}$ & Mass & \vfront & $\left| v_{\rm flow} \right|$ &  \vrad & \vrot & $crit$ & $\left| \frac{v_{\rm front}}{v_{\rm flow}} \right|$ \\ %$\left| \frac{v_{\rm rad}}{v_{\rm rot}} \right|$
		  && [\msun] &  [\msun] & [\kms] & [\kms] & [\kms] &[\kms] \\
	         \hline                  
	         S1F1T140 & 47 & 194.16 & 4.13 (8.83) & 0.75 (0.175) & 0.36 (0.25) & -0.13 (0.008) & -0.005 (0.084) & 0.22 (0.23) & 6.22 (11.92) \\ %& 44.32 (195.9)
	         S2F1T180 & 28 & 265.55 & 9.48 (9.04) & 0.82 (0.26) & 0.36 (0.22) &-0.11 (0.018) & 0.014 (0.071) & 1.04 (1.70) & 5.39 (13.74) \\ %& 6.34 (15.96)
	         S1F2T140 & 38 & 196.17 &5.16 (7.05) & 1.25 (0.39) & 0.49 (0.42) &-0.11 (0.008) & 0.012 (0.088) & 0.37 (0.36) & 7.12 (11.52) \\%& 5.67 (10.46)
	         S2F2T100 & 35 & 63.18 &1.81 (2.18) & 1.29 (0.67) & 0.37 (0.42) & -0.07 (0.011) & 0.003 (0.069) & 0.03 (0.02) & 6.75 (11.50) \\%& 3.17 (3.87)
		 \hline
	         \hline
		\end{tabular}

	\caption{ The mean value and its standard deviation (in parenthesis) of the mass and velocities of all the sub-filaments in the simulated regions. $N_{\rm fil}$ is the number of sub-filaments, M$_{{\rm tot}}$ the total mass in sub-filaments in the network, and $crit$ is the critical line mass \textbf{ratio} in Equation \ref{eq:crit}. The mean absolute magnitude of \vflow is shown so that it can be fairly compared to \vfront. The ratio $|v_{\rm front} / v_{\rm flow}|$ is found by first calculating the ratio for each individual sub-filament and then taking the average. 
	\label{summary_table}
		}
\end{table*}
%$\Delta$\vflow & 0.81 (3.67), -0.30 (4.40), -1.72 (9.51), 2.80 (6.84)

\begin{figure}
\begin{center}
%\begin{tabular}{c c c}
\includegraphics[width=2.4in]{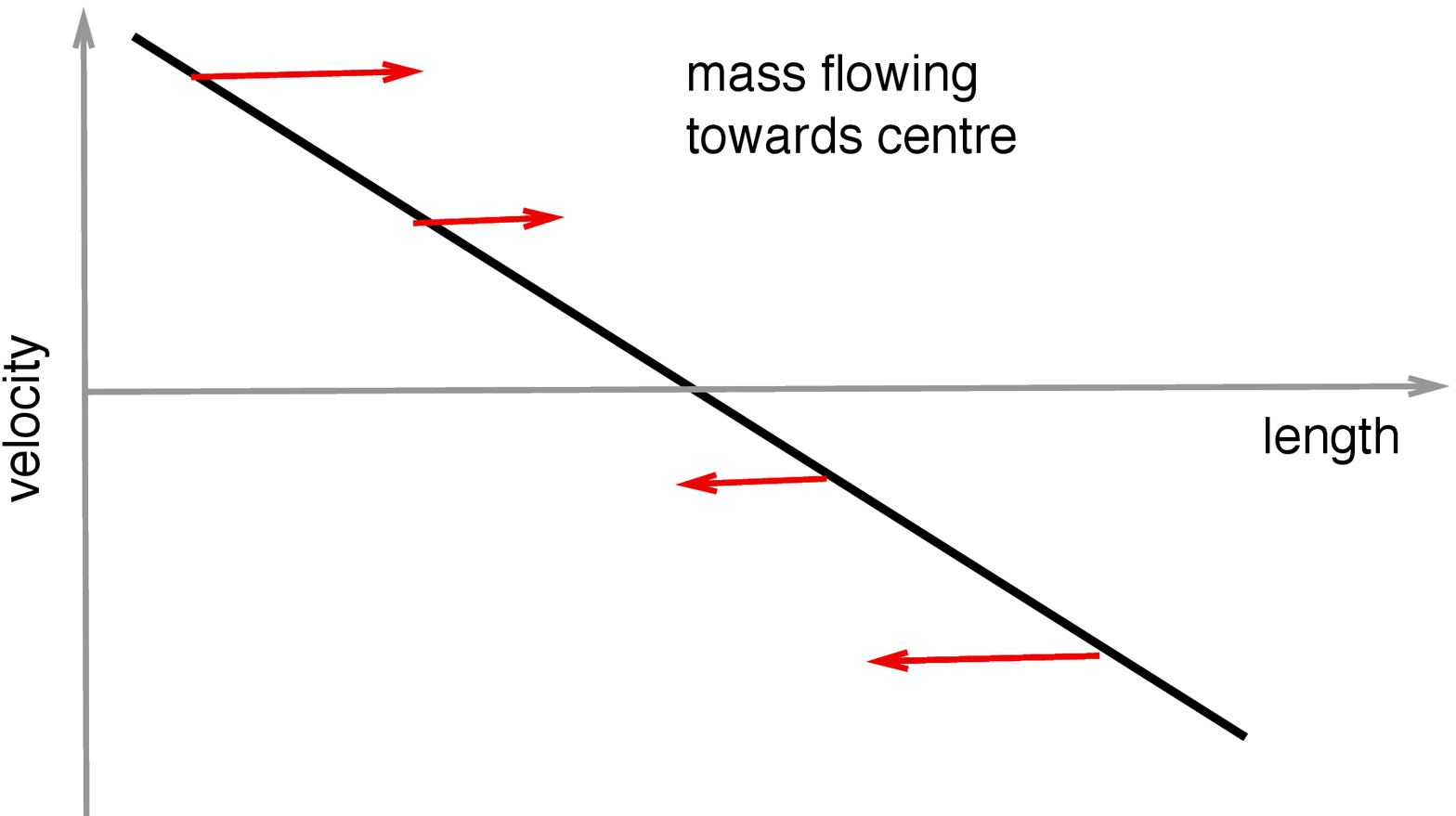}
\includegraphics[width=2.4in]{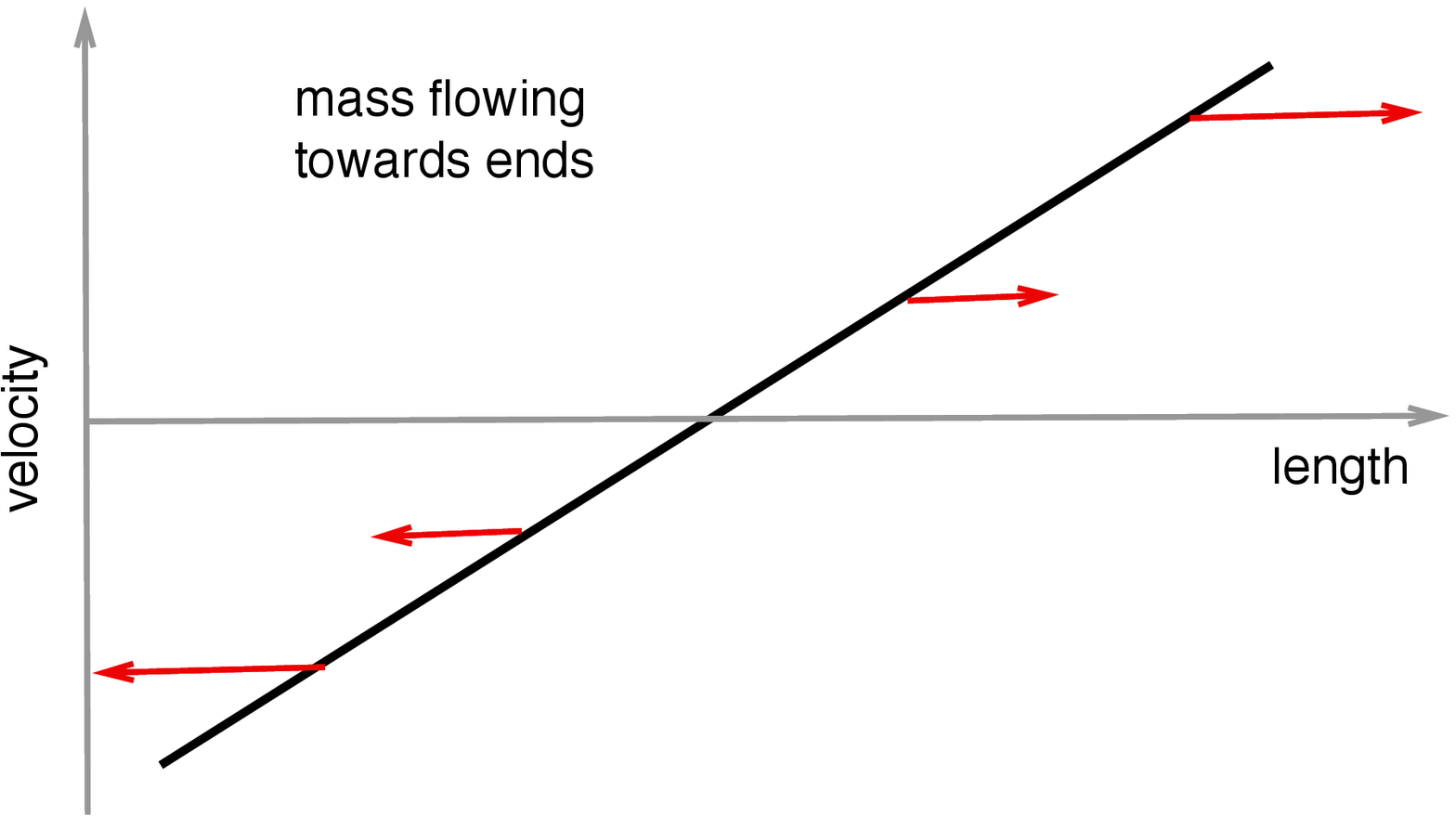}
%\end{tabular}
\caption{A simple schematic showing how gradients in the velocity along the sub-filaments spine correspond to mass flows.}
\label{flow_cartoon}
\end{center}
\end{figure}

\begin{figure*}
\begin{center}
\begin{tabular}{c c}
\begin{overpic}[scale=0.45]{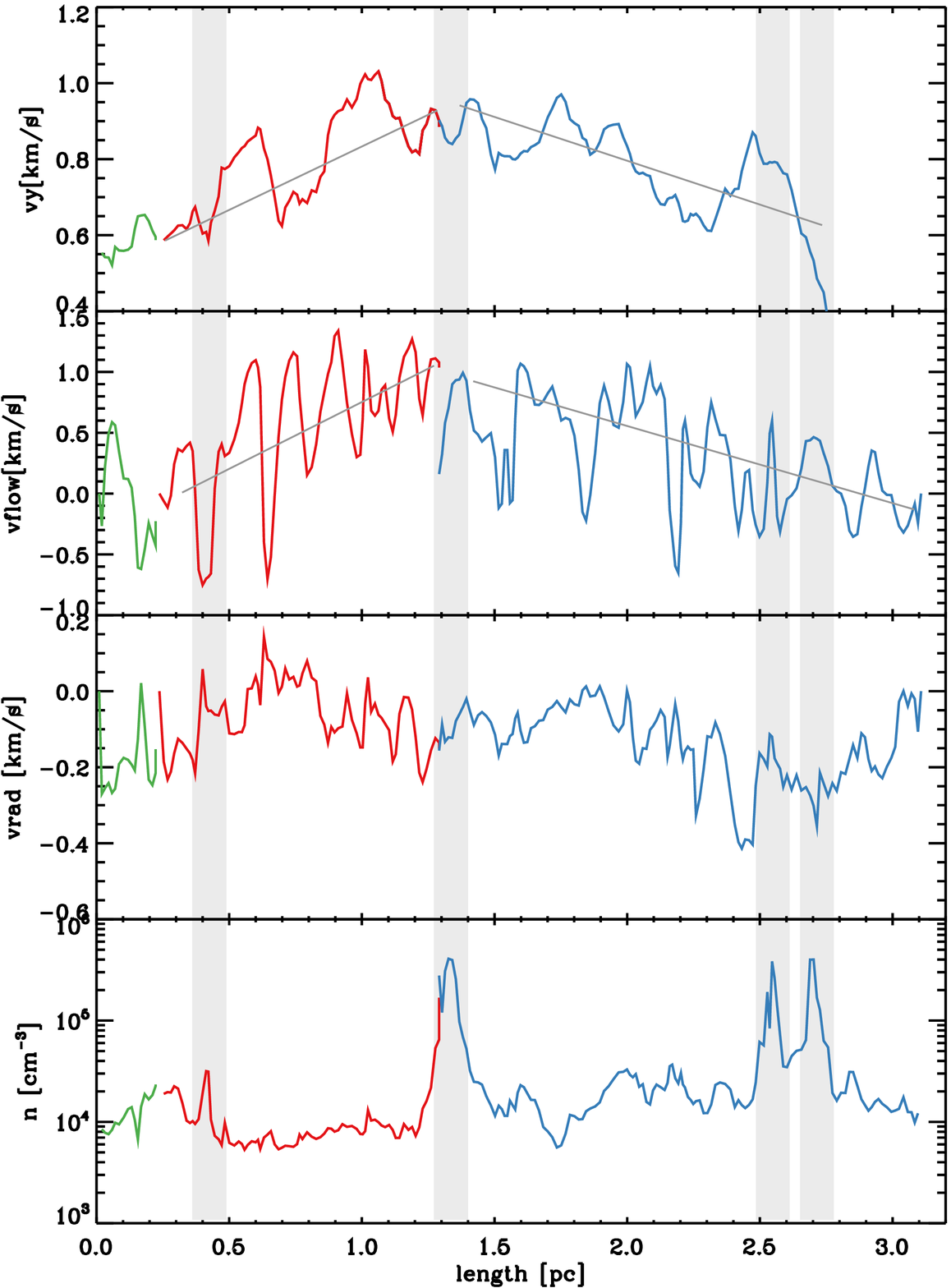}
\put (18,95) {\makebox(0,0){{\bf S1F1T303}}}
\end{overpic}
\begin{overpic}[scale=0.45]{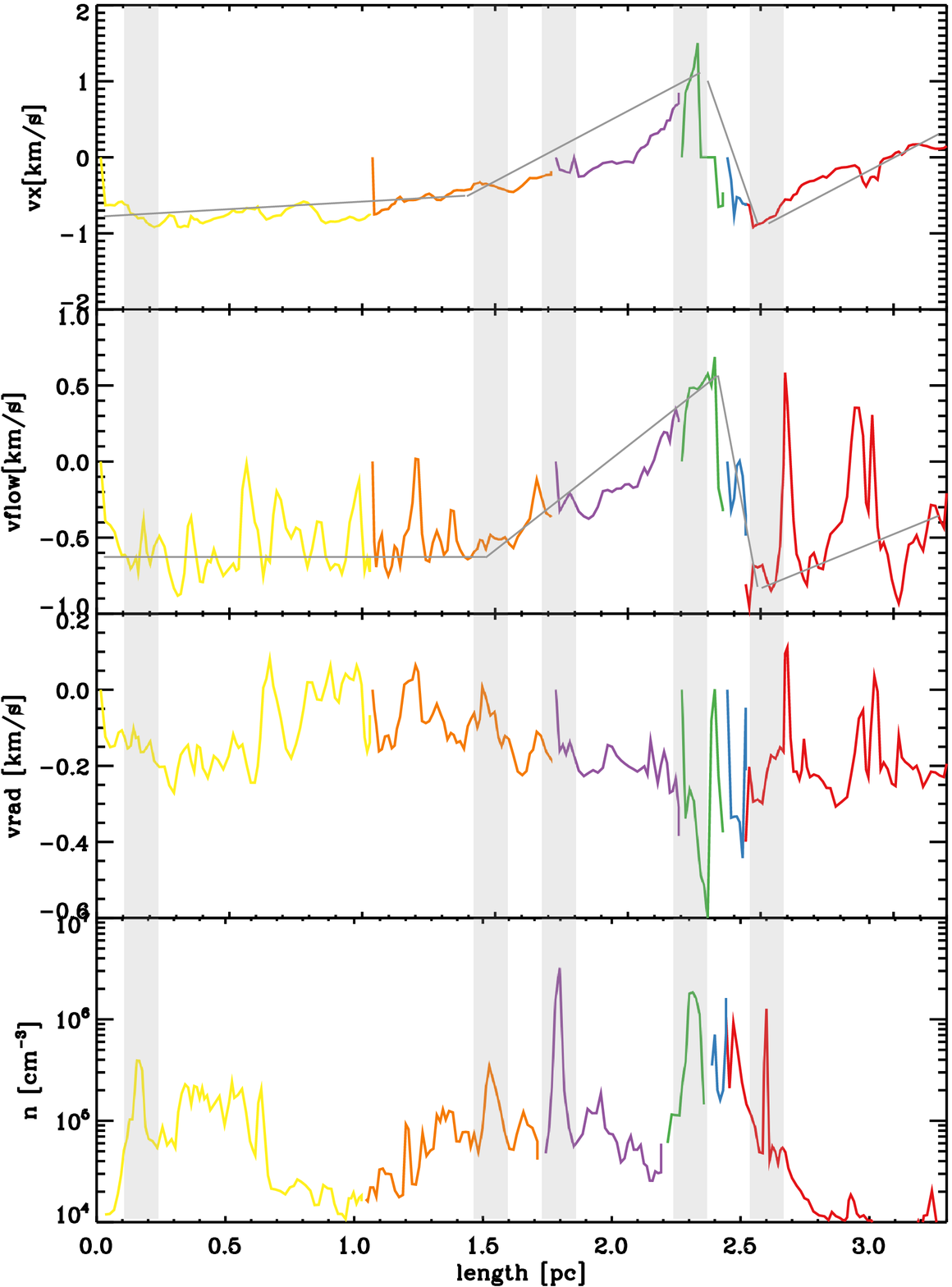}
\put (18,95) {\makebox(0,0){{\bf S2F1T390}}}
\end{overpic}
\end{tabular}
\caption{The velocity along one line-of-sight, \vflow, \vrad and density along one axis for a series of connected sub-filaments. Each sub-filament is shown by a different line colour. The grey shading highlight the positions of peaks in the density profile along the spine, which correspond to the locations of dense cores, and the thin grey line shows the trend of the velocities. There are oscillations in the velocity field along the filament spine, but generally there is both radial contraction and a flow of material along the sub-filaments onto the cores.}
\label{flow_spine}
\end{center}
\end{figure*}

If there are velocity gradients along the sub-filament spines, there must be a net flow of mass along the sub-filaments. In Figure \ref{flow_cartoon} we illustrate this using a simple cartoon. Figure \ref{flow_spine} shows the variation in the sub-filament velocities along their spine. We plot a sample of connected sub-filaments, which are located in the black boxes shown in Figure \ref{columns}. These sub-filaments join smoothly into one another, but are classed as separate objects by \disperse as there are sharp changes in the angle between their boundaries. Each sub-filament is shown in a different colour. The lowest panel of the plot shows the density along the filament spines. The peaks in the density distribution represent the location of dense gas cores. To help us to study the velocities in the vicinity of these cores, we have highlighted their positions with a grey background.

The top panel in Figure \ref{flow_spine} shows the $y$-velocity along the sub-filaments from S1F1T303, and the $x$-velocity for sub-filaments in S2F1T390. We chose to plot $x$- or $y$-velocities depending on the orientation of the box in Figure \ref{columns}. The second panel shows the decomposed component \vflow along the filament spine. In both cases there are oscillations along the spines (the two cases are not identical because the sub-filaments are not completely straight but have kinks and twists). Such oscillatory motions have been previously observed by \citet{Hacar11} where they correlated well with the locations of dense cores. In subsequent studies, however, oscillations were still seen, but they no longer correlated with the cores \citep{Tafalla15}. In our simulated sub-filaments the oscillations and cores are not well correlated, implying that they are transient motions that vary depending on the random turbulent field in which the sub-filaments are embedded. We also calculated the Fourier transform of \vflow along the filament spines and found no dominant frequency mode. Again, this suggests that the oscillations are driven by the surrounding turbulent field, rather than being a consequence of a periodic gravitational instability along the filament spine.

In addition to the oscillations, there are also larger scale trends in the velocity field, shown by the thin grey lines in Figure \ref{flow_spine}. For example, in the sub-filament in S1F1T303 shown in red, there is a net positive gradient along the sub-filament. By examining the cartoon shown in Figure \ref{flow_cartoon} we can see that this corresponds to a net flow of mass along the sub-filament towards a core at the end. The adjoining blue sub-filament has a negative net gradient along its length, which corresponds to a mass flow towards the centre. It is therefore clear that mass can be transferred along the sub-filaments. The cores tend to appear preferentially where the large-scale velocity field changes and where there is a flow of mass towards the core. This  demonstrates the importance of gravity in forming the cores. Depending on the relative strength of the tidal field compared to the self-gravity of the sub-filament, the sub-filament can be interpreted either as an accretion flow that feeds an existing over-density at the end of the sub-filament, or instead it may fragment to form its own cores. We will discuss sub-filament fragmentation further in Section \ref{results2}.

\begin{figure}
\begin{center}
\includegraphics[width=3in]{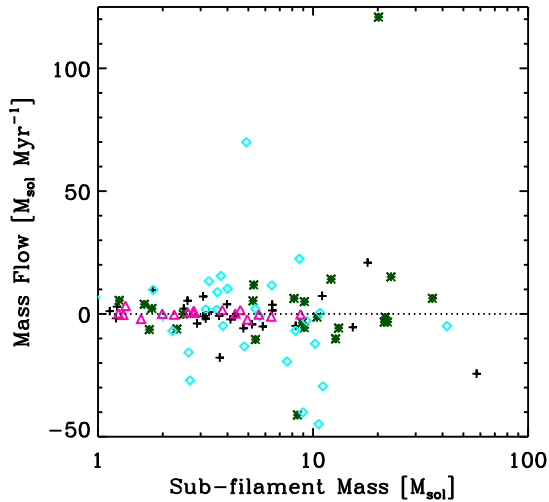}
\caption{The net mass flow along individual sub-filament spines. Black crosses show sub-filaments in S1F1, green asterisks show sub-filaments in S2F1, cyan diamonds show sub-filaments in S1F2, and pink triangles show sub-filaments in S2F2.}
\label{Mflow}
\end{center}
\end{figure}

To estimate the magnitude of such flows we cannot just sum the mass moving along the sub-filaments, because if we were to take the scenario in Figure \ref{flow_cartoon} the net total would come to zero in both cases despite there being strong mass flows. Instead we fit a straight line to the flow velocity along every sub-filament spine weighted by the mass in each segment for each of the four simulations shown in Figure \ref{columns}. The gradient of this line is then used to obtain a crude order-of-magnitude estimate of the net flow of mass per parsec using the relationship $\dot{M}_{\rm flow}=\Delta v_{\rm  flow} M_{\rm fil}$ where $M_{\rm fil}$ is the total mass of the filament. Figure \ref{Mflow} shows the estimated value of $\dot{M}_{\rm flow}$ for each sub-filament in the four simulations, plotted against its mass. 

The mass flows are both negative and positive, meaning that in some sub-filaments mass is flowing towards the centre, and in others to the end. For the dense sub-filaments, tens of solar masses of gas can be moved along the sub-filament per Myr. For comparison in the super-critical, southern filament of Serpens South, \cite{Kirk13} measured a gas flow of $30 \: {\rm M_{\odot} \: Myr^{-1}}$  along the filament, which is of the same order of magnitude as the values we measure in our simulations. Large-scale filaments have also been observed to have converging motions attributed to collapse along their long axis \citep[e.g][]{Zernickel13,Peretto14}. However, it is important to stress that this process can be both constructive and destructive, as sub-filaments with a positive $\dot{M}_{\rm flow}$ will become increasingly diffuse at their centre. The arrow of time does not point inevitably towards increasing densities, and many sub-filaments will never form stars. Instead they may either be sheared apart by turbulence, or cannibalised by nearby accreting cores.

\subsection{Radial and Rotational Velocities}
\label{sect:rad_rot}

\begin{figure*}
\begin{center}
\begin{tabular}{c c c}
\begin{overpic}[scale=0.4]{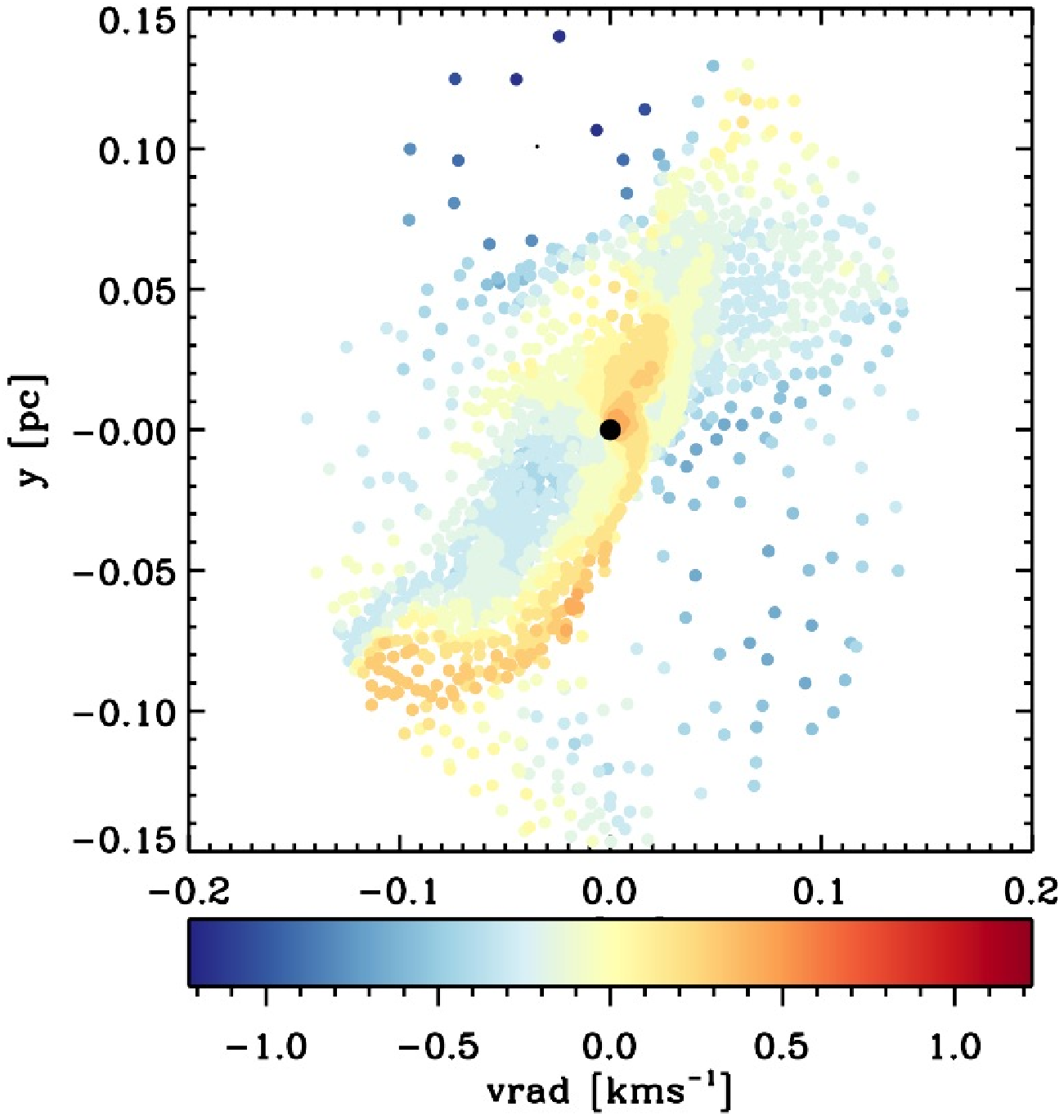}
%\put (35,90) {\makebox(0,0){{\bf crit = 0.75}}}
\end{overpic}
\begin{overpic}[scale=0.4]{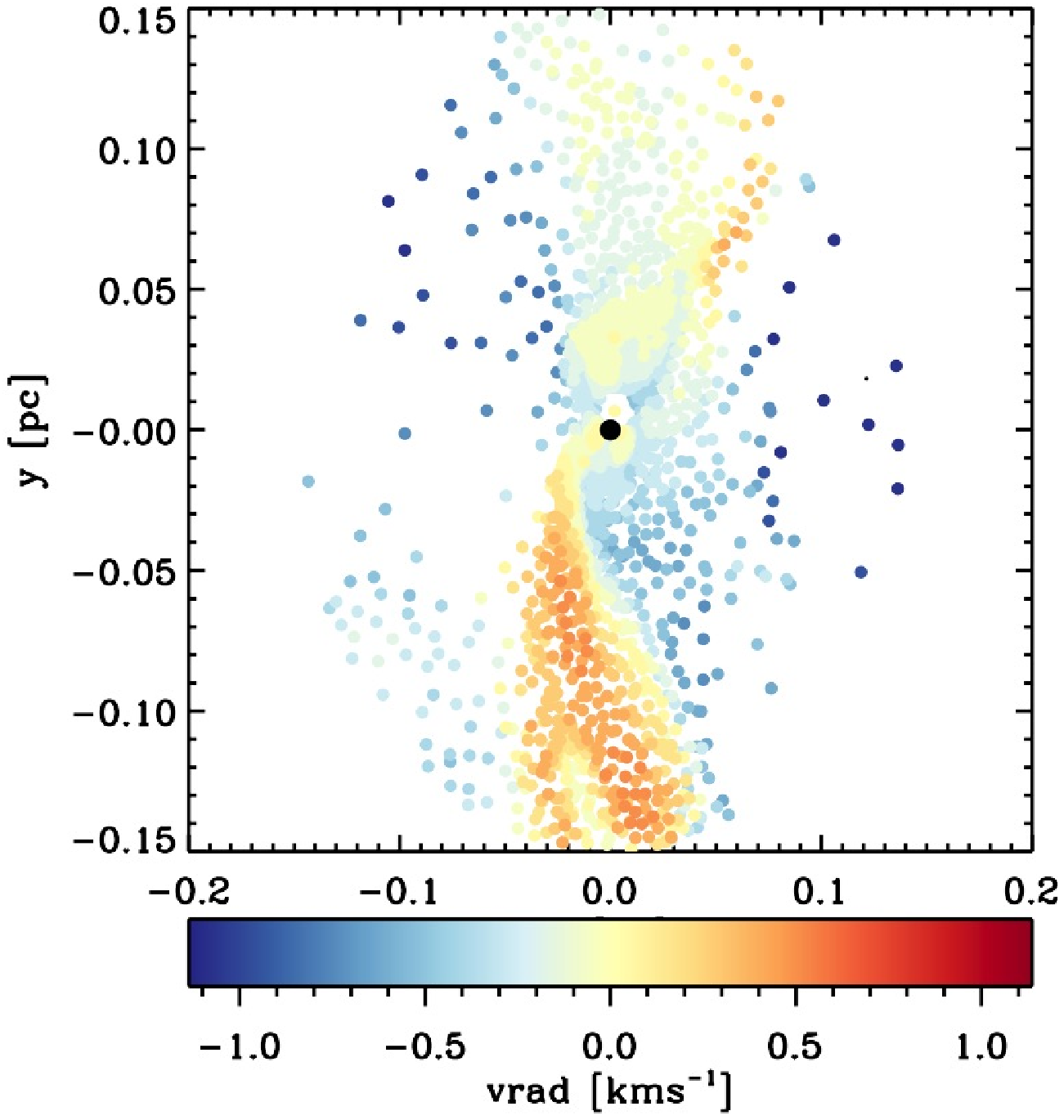}
%\put (35,90) {\makebox(0,0){{\bf crit = 1.46}}}
\end{overpic}
\begin{overpic}[scale=0.4]{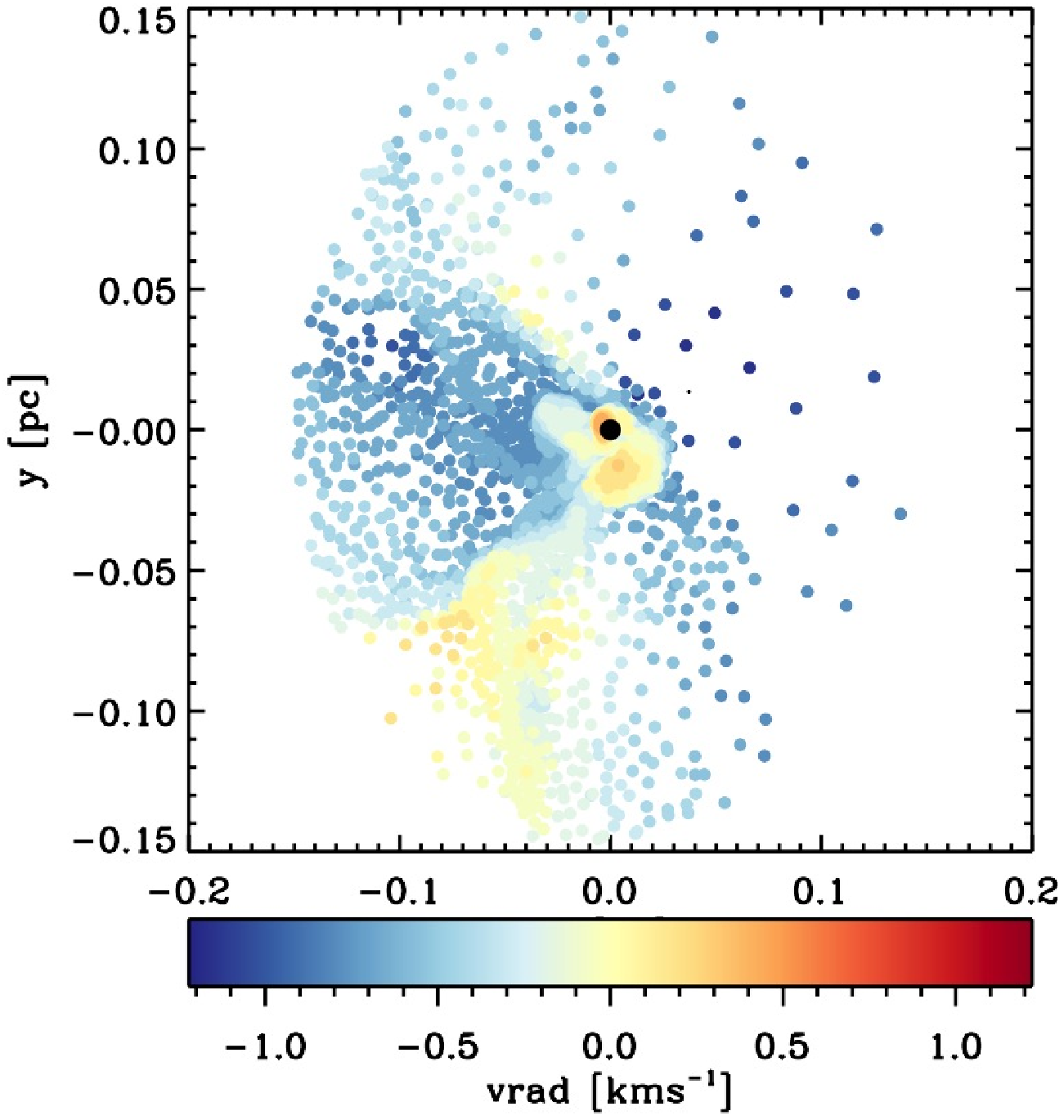}
%\put (35,90) {\makebox(0,0){{\bf crit = 1.35}}}
\end{overpic}
\end{tabular}
\caption{Illustrative slices showing the radial velocity in some sub-filament cross-sections. Red points show the position of cells with a net outward velocity and blue shows cells moving inwards.}
\label{slice_rad}
\end{center}
\end{figure*}

\begin{figure*}
\begin{center}
\begin{tabular}{c c c}
\begin{overpic}[scale=0.4]{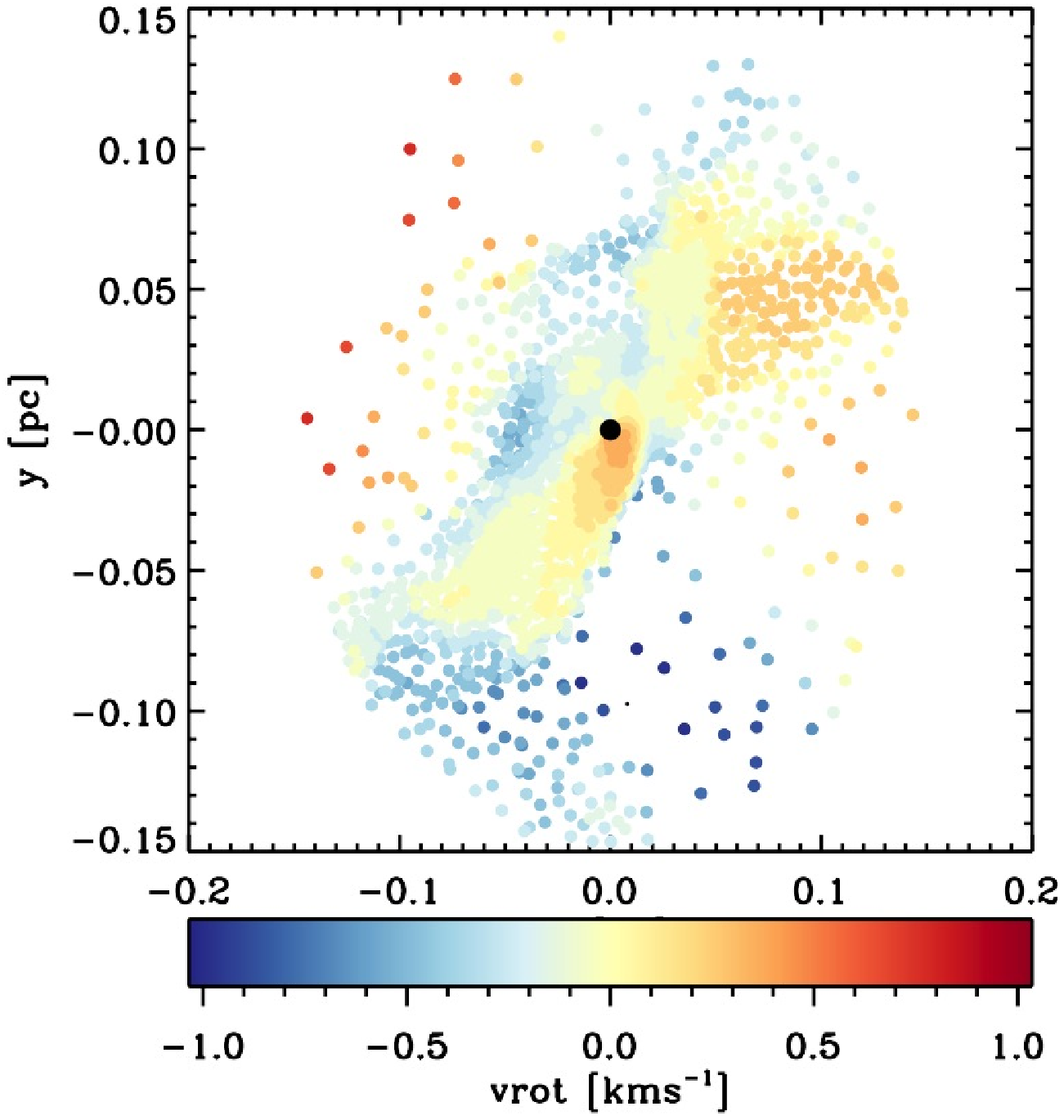}
%\put (35,90) {\makebox(0,0){{\bf crit = 0.75}}}
\end{overpic}
\begin{overpic}[scale=0.4]{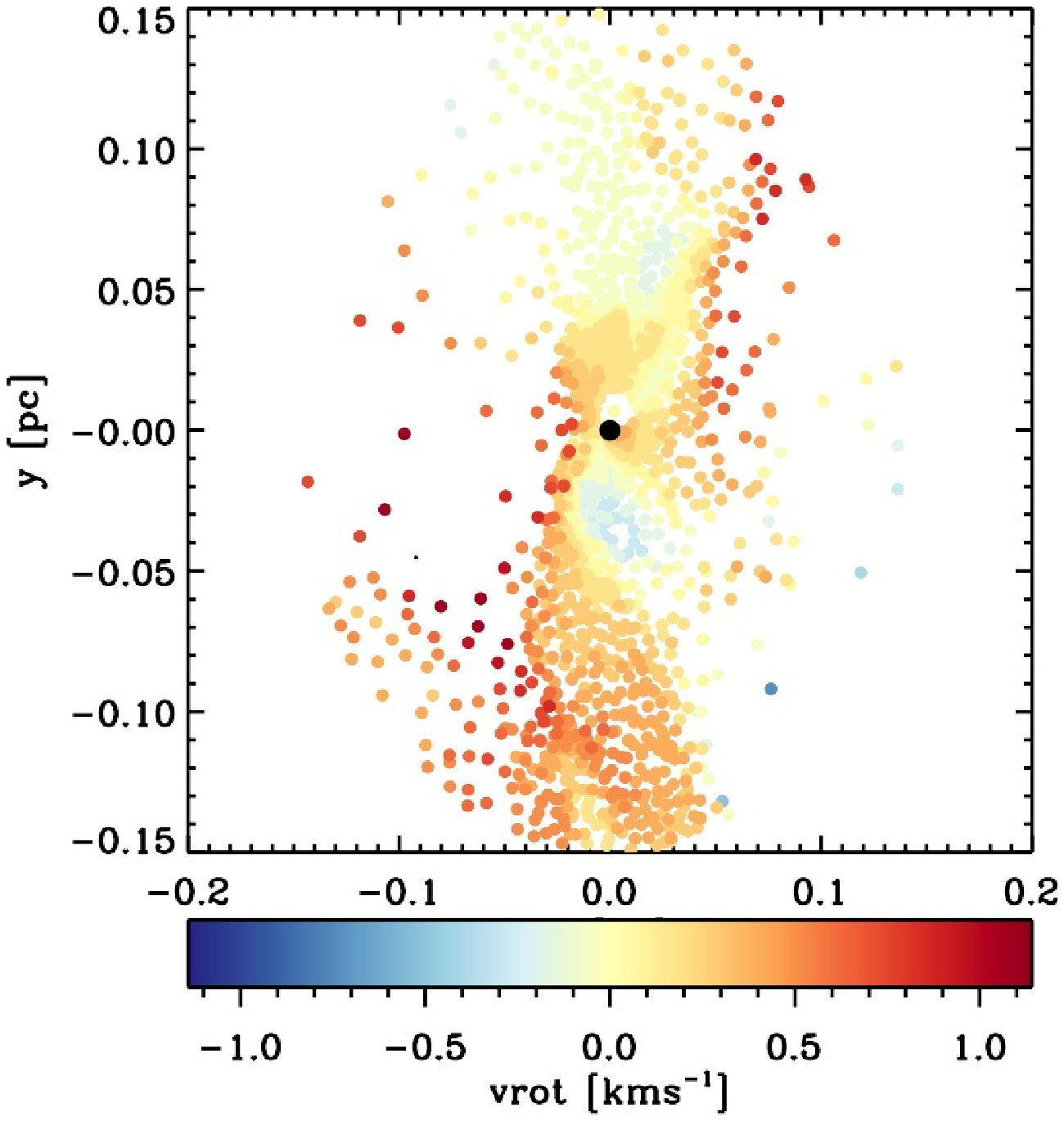}
%\put (35,90) {\makebox(0,0){{\bf crit = 1.46}}}
\end{overpic}
\begin{overpic}[scale=0.4]{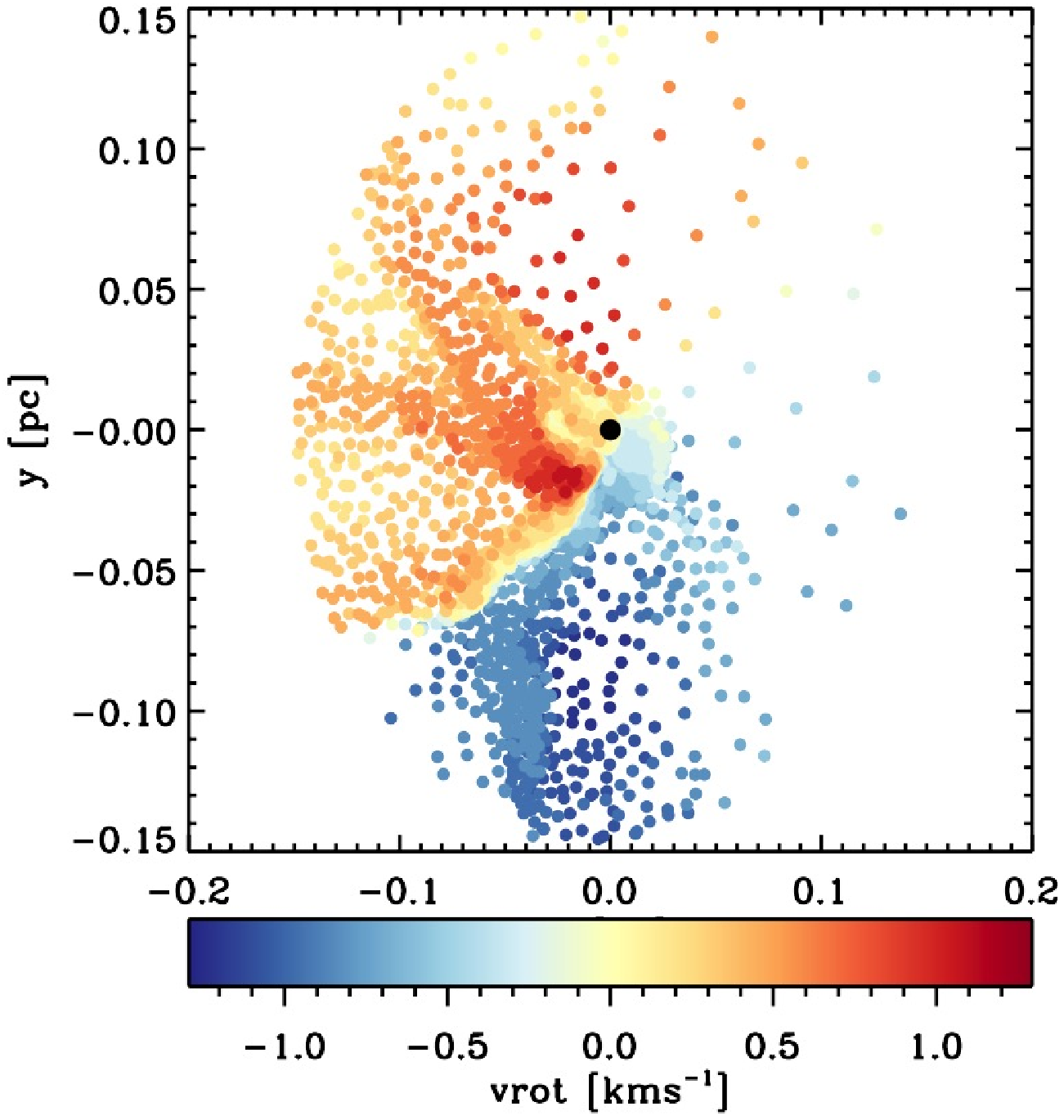}
%\put (35,90) {\makebox(0,0){{\bf crit = 1.35}}}
\end{overpic}
\end{tabular}
\caption{Illustrative slices showing the rotational velocity in some sub-filament cross-sections. Red points show the position of cells with an anti-clockwise velocity and blue shows cells moving clockwise. }
\label{slice_rot}
\end{center}
\end{figure*}

Having investigated the net velocities along the sub-filament spines, we now turn our attention to the local residual velocities. The radial and rotational velocities were obtained by subtracting off the centre-of-mass velocity for each sub-filament segment and then decomposing the velocities as discussed in Section \ref{meth:fil}. In Figure \ref{flow_spine} we also plot the mean radial velocity for each segment along the sub-filament spines. In almost all cases the radial velocity is between 0 and -0.2 \kms~indicating subsonic radial contraction of gas onto the spine. The filaments are, therefore, accreting radially as well as along the filament spine. As before, the radial contraction is not uniform, but instead varies along the sub-filament.

\begin{figure}
\begin{center}
\includegraphics[width=3in]{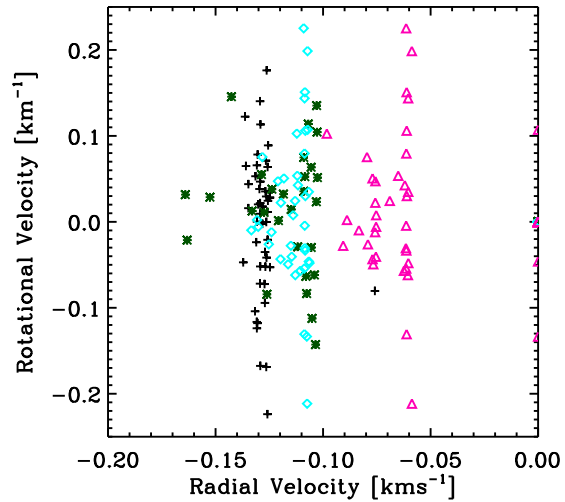}
\caption{The mean radial vs rotational velocity for all the sub-filaments. Black crosses show sub-filaments from S1F1, green asterisks show sub-filaments from S2F1, cyan diamonds show sub-filaments from S1F2, and pink triangles show sub-filaments from S2F2. }
\label{rad_rot}
\end{center}
\end{figure}

To understand the details of the residual motions we show in Figures \ref{slice_rad} and \ref{slice_rot} the radial and rotational velocities from three sub-filament segments in S1F1 and S2F2 as illustrative examples. The projection is face-on so that the spine vector points into the page. Figure \ref{slice_rad} shows the radial velocity of the gas cells in the segments. In each case there is a high velocity flow of the more diffuse material onto the sub-filament spine centre, but at the centre the velocities are more quiescent. As discussed in \citeauthor{Smith14b}~(2014b), the sub-filament sections are elongated, not circular, which shows that the sub-filaments are not true cylinders. The flow of gas onto the sub-filaments is more reminiscent of a colliding flow than of a radial inflow and there is still some gas moving away from the sub-filament centre. Figure \ref{slice_rot} shows the rotational velocity of the gas cells in the segments. There is no strong evidence for uniform rotation about the spine centre. While there are velocities perpendicular to the radial direction these are not ordered at scales of $\sim0.1$ pc (roughly the sonic scale) and seem to be from random turbulence or shearing in the field rather than rotation.

Figure \ref{rad_rot} shows the mass-weighted mean radial and rotational velocities of all the sub-filaments. For each simulation the radial velocities tend to cluster around the same value, since the sub-filaments are embedded in the same turbulent bulk flow. For both \vrad and \vrot, the velocities are subsonic despite being measured over a $r=0.15$~pc region. This is due to most of the mass being concentrated at the sub-filament centre where it can be supported by thermal motions (a volume-weighted mean does indeed give a larger velocity). A similar result was found in \citet{Smith12a}, where the velocity dispersion measured in simulated N$_2$H$^{+}$ emission from cores embedded in filaments was narrower than that calculated by a volume-weighted average. We will discuss the observational consequences of the coherency of the sub-filaments velocities in Section \ref{discussion} and in future papers.
%While these are only illustrative examples, we include in the online material \textit{(at insert url)} movies showing the segments along the entire sub-filament for a range of sub-filaments with different zoom-ins. 

% here I have forced latex to make a single landscape page
\begin{landscape}

\thispagestyle{plain}

\begin{figure}
\begin{center}
\begin{tabular}{c c c c}
\begin{overpic}[scale=0.32]{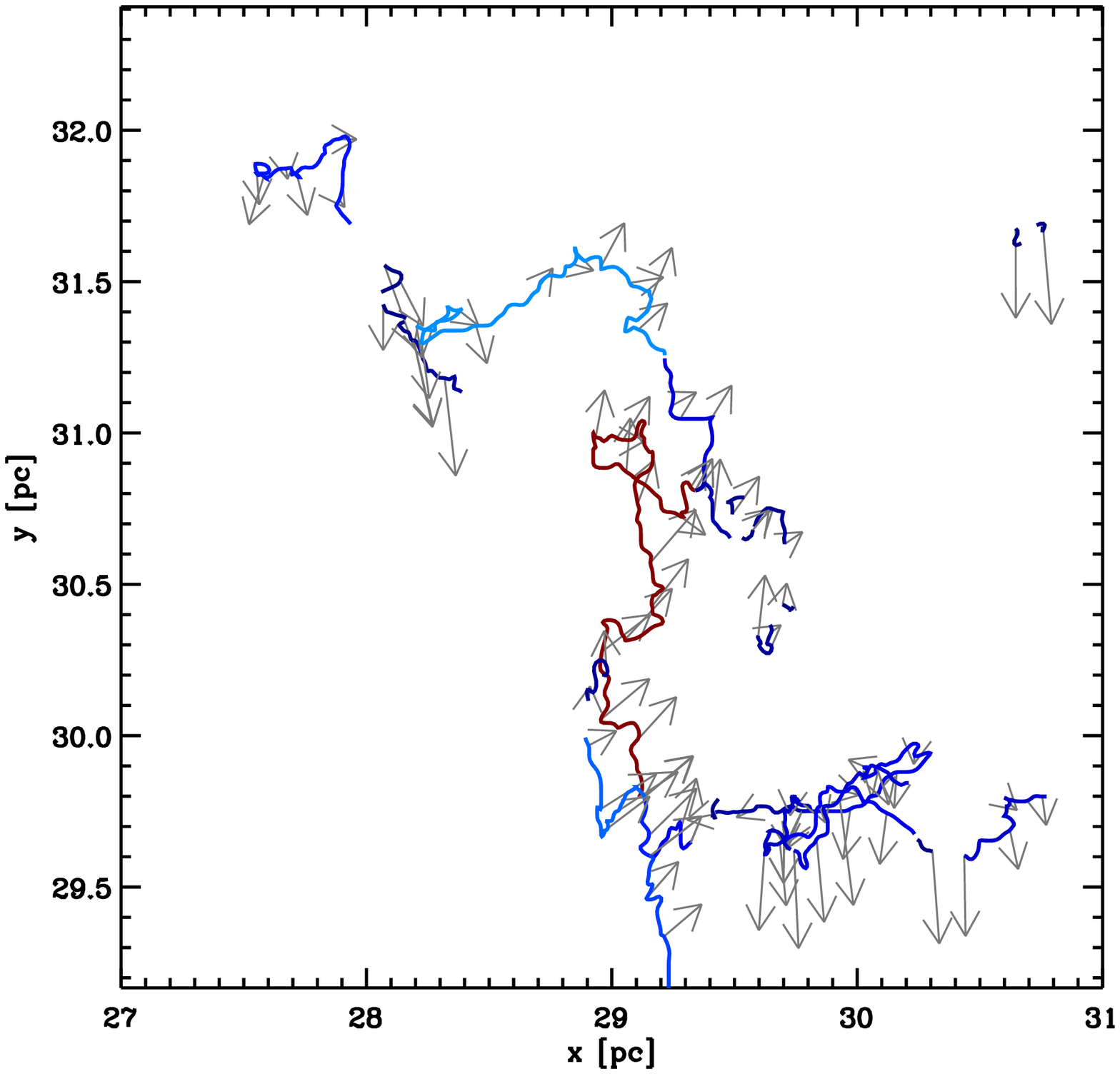}
\put (75,90) {\makebox(0,0){{\bf S1F1T260}}}
\put (75,85) {\makebox(0,0){{\bf N$_{{\rm fil}}$ = 31}}}
\put (75,80) {\makebox(0,0){{\bf N$_{{\rm sink}}$= 0}}}
\end{overpic}
\begin{overpic}[scale=0.32]{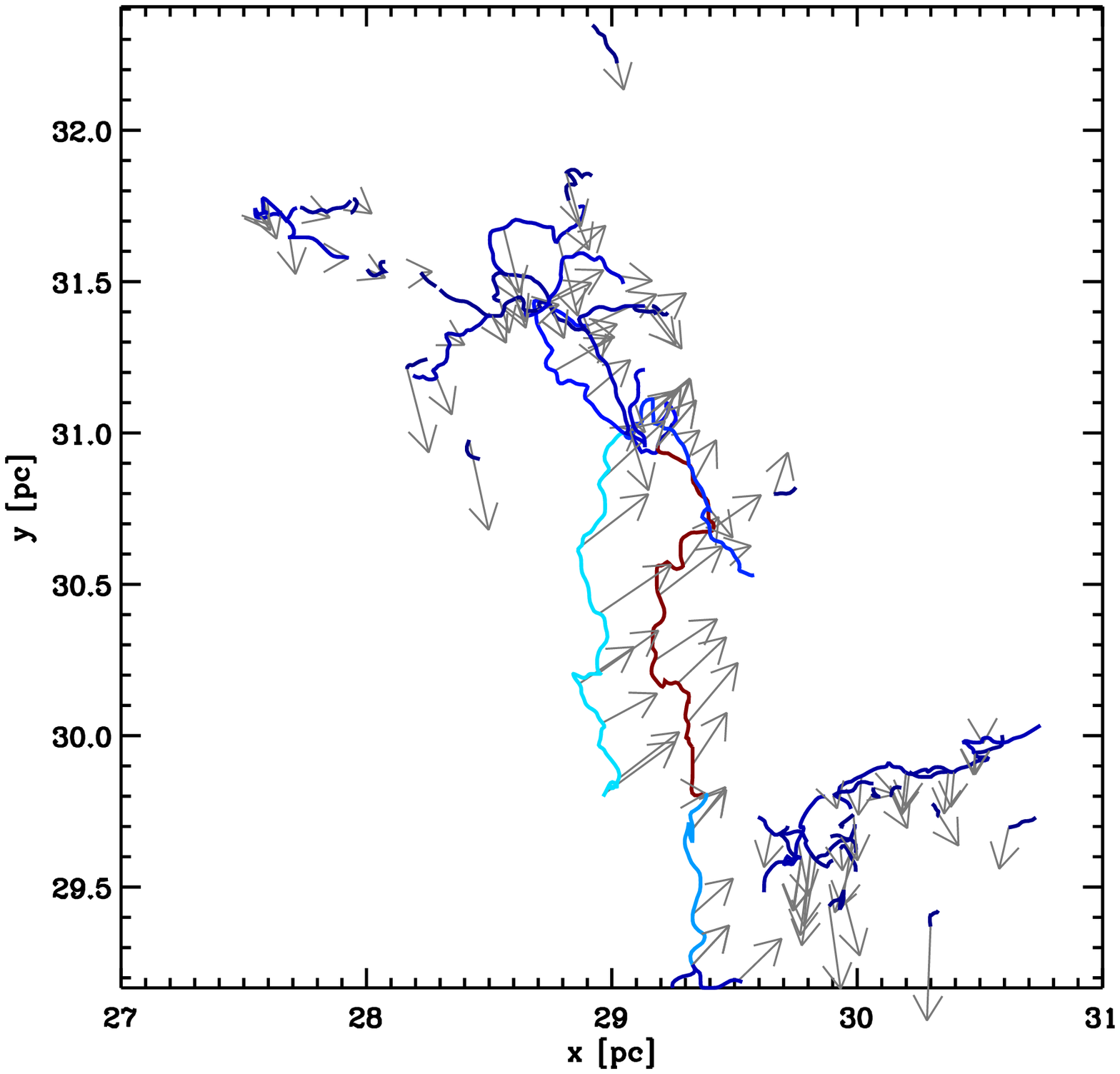}
\put (75,90) {\makebox(0,0){{\bf S1F1T281}}}
\put (75,85) {\makebox(0,0){{\bf N$_{{\rm fil}}$ = 61}}}
\put (75,80) {\makebox(0,0){{\bf N$_{{\rm sink}}$= 0}}}
\end{overpic}
\begin{overpic}[scale=0.32]{./figs_final/S1F1T140Filament_map_vel}
\put (75,90) {\makebox(0,0){{\bf S1F1T303}}}
\put (75,85) {\makebox(0,0){{\bf N$_{{\rm fil}}$ =47}}}
\put (75,80) {\makebox(0,0){{\bf N$_{{\rm sink}}$= 1}}}
\end{overpic}
\begin{overpic}[scale=0.32]{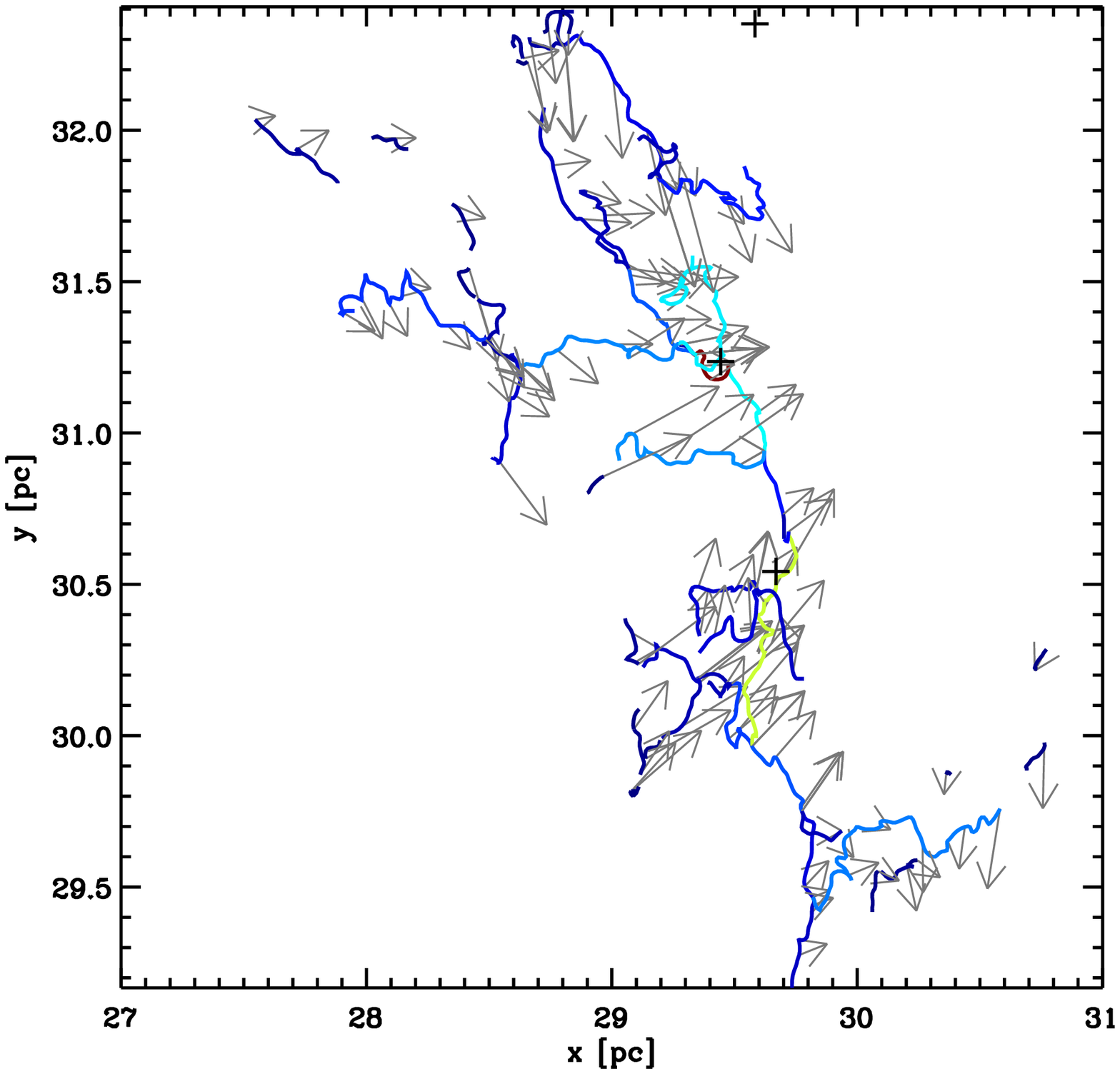}
\put (75,90) {\makebox(0,0){{\bf S1F1T325}}}
\put (75,85) {\makebox(0,0){{\bf N$_{{\rm fil}}$ = 56}}}
\put (75,80) {\makebox(0,0){{\bf N$_{{\rm sink}}$= 2}}}
\end{overpic}\\
\begin{overpic}[scale=0.32]{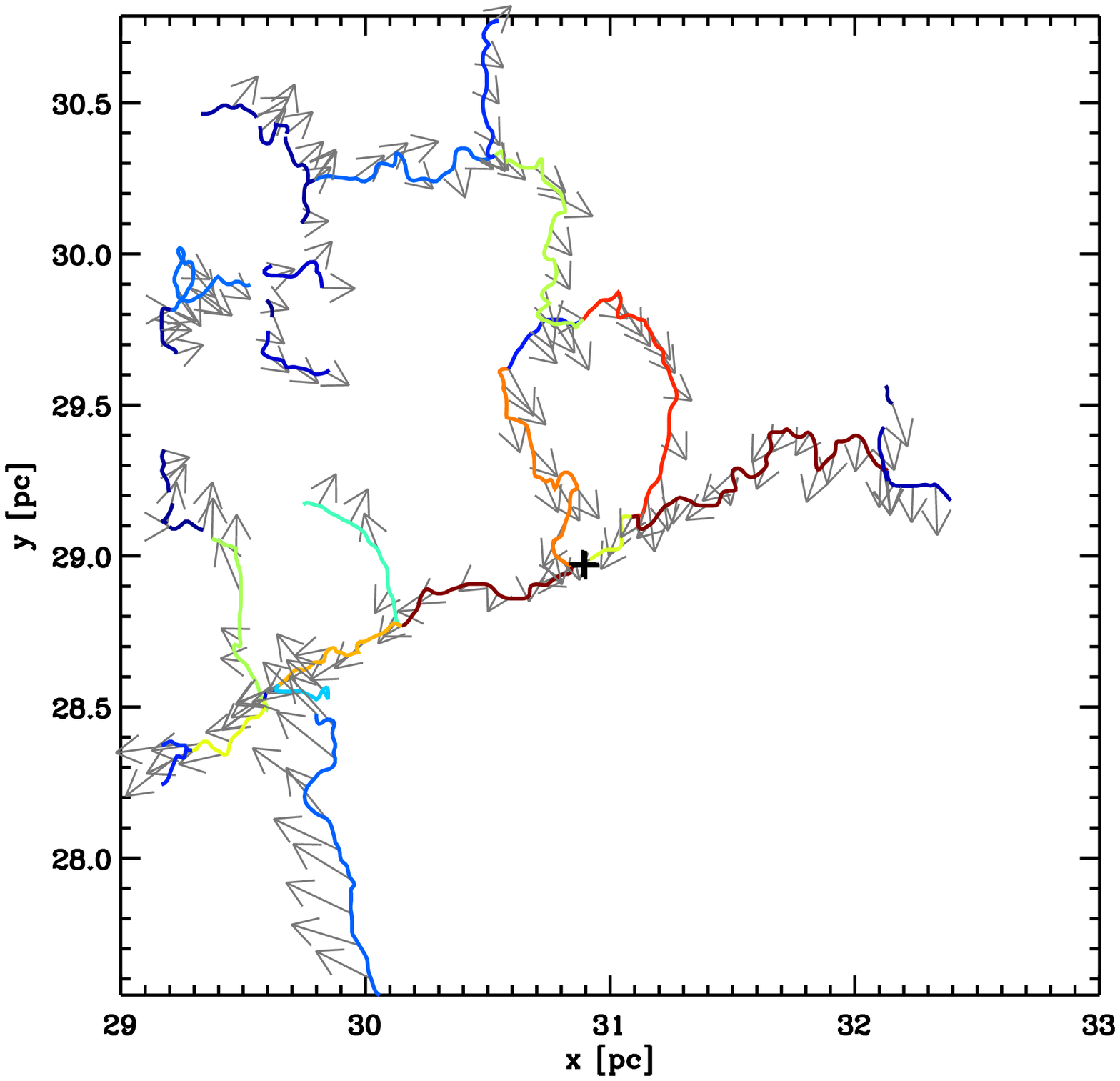}
\put (75,90) {\makebox(0,0){{\bf S2F1T368}}}
\put (75,85) {\makebox(0,0){{\bf N$_{{\rm fil}}$ = 34}}}
\put (75,80) {\makebox(0,0){{\bf N$_{sink}$= 6}}}
\end{overpic}
\begin{overpic}[scale=0.32]{./figs_final/S2F1T180Filament_map_vel}
\put (75,90) {\makebox(0,0){{\bf S2F1T390}}}
\put (75,85) {\makebox(0,0){{\bf N$_{{\rm fil}}$ = 28}}}
\put (75,80) {\makebox(0,0){{\bf N$_{sink}$= 9}}}
\end{overpic}
\begin{overpic}[scale=0.32]{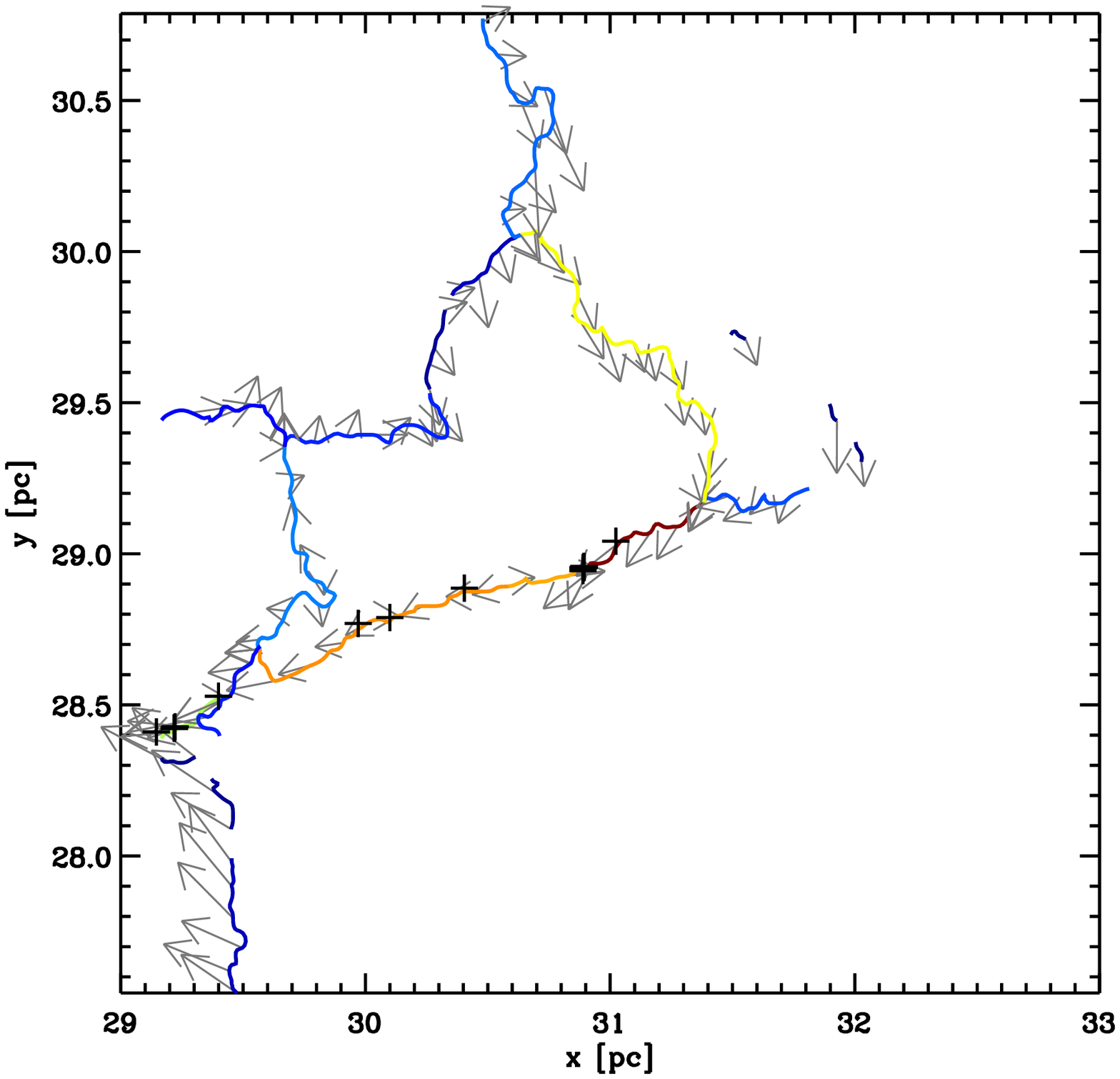}
\put (75,90) {\makebox(0,0){{\bf S2F1T411}}}
\put (75,85) {\makebox(0,0){{\bf N$_{{\rm fil}}$ = 20}}}
\put (75,80) {\makebox(0,0){{\bf N$_{{\rm sink}}$= 16}}}
\end{overpic}
\begin{overpic}[scale=0.32]{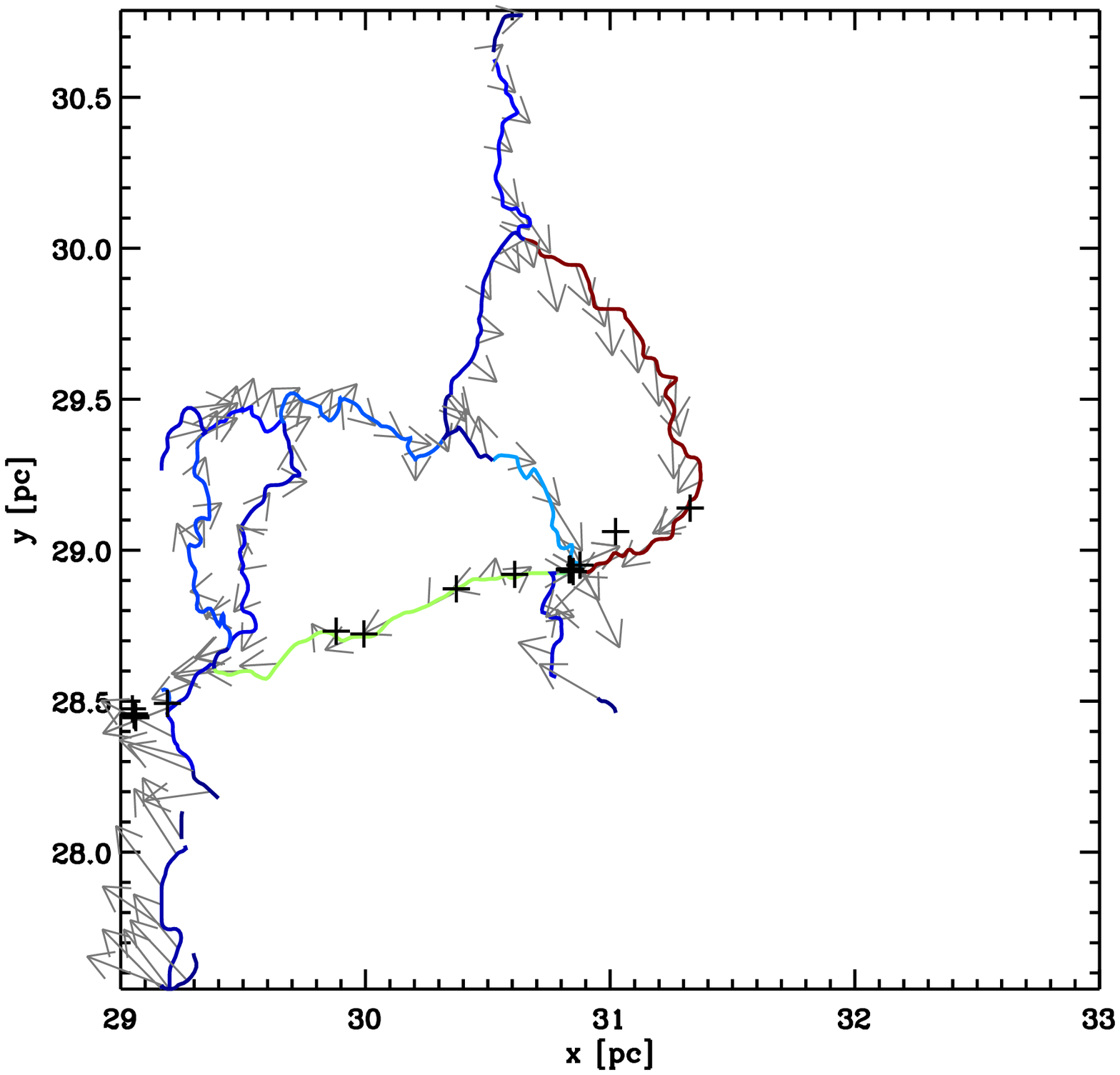}
\put (75,90) {\makebox(0,0){{\bf S2F1T433}}}
\put (75,85) {\makebox(0,0){{\bf N$_{{\rm fil}}$ = 26}}}
\put (75,80) {\makebox(0,0){{\bf N$_{{\rm sink}}$= 17}}}
\end{overpic}
\end{tabular}
\caption{The evolution of S1F1 and S2F2 at intervals of $2.17\E^5$ years (ten snapshot intervals) as seen in the $x$-$y$ plane. The colours denote the mass in each sub-filament, with dark blue being the least massive and red the most massive. Sink particles are shown by crosses. The grey arrows show the direction of the filament velocity at every tenth filament spine point. While the overall shape of the network stays similar the number and positions of the sub-filaments do change over time.}
\label{Evolution}
\end{center}
\end{figure}

\end{landscape}

\section{Time Evolution and Fragmentation}\label{results2}

The filamentary networks presented in the last section were studied at a single point in time. However, in order to investigate how the sub-filaments are involved in star formation, we need to consider their dynamical evolution. Figure \ref{Evolution} shows the evolution of the filament network in S1F1 and S2F1. While the overall outline of the network stays similar, the number of sub-filaments and their positions vary.

In Table \ref{evo-Nfil} we show how the masses of the sub-filaments and the number of sink particles evolve. The time quoted for each snapshot is measured from the beginning of the simulation. The number of sub-filaments shows no consistent trend with time, which emphasises that filamentary substructure is a \textit{general} feature of molecular clouds. The mass in the sub-filaments, on the other hand, does clearly increase as the cloud evolves. This is because the mass in the region is increasing due to global converging motions. The converging motions were initially the result of the large-scale turbulent modes gathering gas together, but over time these regions become gravitationally unstable, causing the gas to collapse and the density to increase. Over the studied period the mass in the sub-filament spines in S1F1 increased by $2.15 \E^{-4} \: {\rm M_{\odot} \: yr^{-1}}$ and in S2F1 by $9.7 \E^{-5} \: {\rm M_{\odot} \: Myr^{-1}}$. Substantial amounts of mass can therefore be gathered into dense coherent sub-filaments over the lifetime of the cloud, making it more susceptible to fragmentation.

\begin{table*}
	\centering
		\begin{tabular}{l c c c c c c c c}
   	         \hline
	         \hline
		  Filament & Time & $N_{\rm fil}$ & Mean mass & Total mass  &  Mean $crit$ & N$_{\rm crit}$ &Max $crit$ & $N_{\rm sink}$ \\
		  & [Myr] & & [\msun] &  [\msun] \\
	         \hline                  
	         S1F1T238 & 2.38 & 30 & 3.82 (3.90) & 114.65 & 0.20 (0.14) & 0 & 0.62 & 0 \\
	         S1F1T260 & 2.60 & 31 & 4.48 (8.40) & 138.86 & 0.18 (0.14)  & 0 & 69 & 0\\
	         S1F1T281 & 2.81 & 61 & 3.06 (6.70) & 186.52 & 0.22 (0.21) & 0 & 0.90 & 0\\
	         S1F1T303 & 3.03 & 47 & 4.13 (8.83) & 194.16 & 0.22 (0.23) & 3 &1.46 & 1\\
	         S1F1T325 & 3.25 & 56 & 5.39 (8.47) & 301.87 & 0.57 (1.34) & 7 & 9.48 & 2\\
		 \hline 
		 S2F1T160 & 3.46 & 33 & 5.77 (8.25) & 190.47 &  0.38 (0.52) & 2 & 2.58 & 1 \\
		 S2F1T170 & 3.68 & 34 & 6.58 (7.28) & 223.71 & 0.53 (0.62) & 5 & 3.65 & 6 \\
		 S2F1T180 & 3.90 & 28 & 9.48 (9.05) & 265.56 & 1.04 (1.70) & 8 & 7.18 & 9 \\
		 S2F1T190 & 4.11 & 20 & 13.01 (15.27) & 260.24 & 1.20 (1.72) & 6 & 5.54 & 16\\
		 S2F1T200 & 4.33 & 26 & 10.61 (15.32) & 275.80 &  1.54 (3.99) & 7 & 20.1 & 17\\
		 \hline
	         \hline
		\end{tabular}
	
	\caption{ The evolution of the sub-filament properties for S1F1 and S2F1. $N_{\rm fil}$ is the number of sub-filaments, $crit$ is the critical line mass in Equation \ref{eq:crit}, $N_{\rm crit}$ is the number of sub-filaments with a critical ratio greater than one, Max $crit$ is the greatest value of $crit$ for the sub-filaments, and $N_{\rm sink}$ is the number of sink particles in the filament. The standard deviation in each mean is shown in brackets to give a feel for the variation in the sample. \label{evo-Nfil}
	}
\end{table*}

The mass per unit length is a useful way of quantifying whether a filament can remain in a stable equilibrium. Using a model of an infinite self-gravitating cylinder with no magnetic support, \citet{Inutsuka97} found that a filament cannot be supported thermally below the critical line mass
\begin{equation}
M_{\rm crit}= 2 c_s^2/G = 16.7\left(\frac{T}{10 \textrm{K}}\right) M_{\odot} \:  {\rm pc}^{-1},
\end{equation}
where $c_s$ is the sound speed, $G$ is the gravitational constant and $T$ is the temperature. We calculate the critical ratio for our filaments using the relation
\begin{equation}
crit=M_{\rm line}/M_{\rm crit} \textcolor{red}{~~,}
\label{eq:crit}
\end{equation}
where $M_{\rm line}$ is the measured mass per pc along the filament spine.

As a consequence of the sub-filament mass increasing, the mean critical ratio of the sub-filaments rises as well, and so does the number of super-critical sub-filaments. In all of the simulated regions, most of the sub-filaments remain sub-critical and do not fragment. However, multiple sink particles (representing star-forming cores) are formed in the sub-filaments that are super-critical. This in agreement with the observations of \cite{Tafalla15} who find that star formation only occurs in the dense `fertile fibres' seen in N$_{2}$H$^{+}$ and not in any of the sub-filaments only detected in C$^{18}$O. We will discuss this further in Section \ref{obs_comp}.

One could get a different estimate of the critical line mass if the internal velocity dispersion were to be added to the sound speed. However, our definition of the sub-filament mass is a very conservative one, as we only include material in the flat part of the density profile in Equation \ref{Plum-dens} to avoid double counting mass where the sub-filaments bend or are in close proximity. Within this region the residual velocities are subsonic as discussed in Section \ref{sect:rad_rot} and so do not contribute significant support. Furthermore, as shown by Figures \ref{flow_spine} and \ref{rad_rot} the radial velocity component is preferentially directed inwards making it unclear how much support such motions offer against collapse. In the simulations we see that cores (i.e. sink particles) are only formed in super-critical sub-filaments, which suggests that our definition is reasonable.

In Section \ref{results1} we discussed the decomposed velocities in the sub-filaments. Table \ref{evo_vel} shows that there is no systematic change in the velocities as the sub-filaments evolve, with the exception of rotation. As shown in Figure \ref{slice_rot}, there is no clear evidence for ordered rotation in the filament, and so the variation in \vrot  probably comes from contributions of random turbulent motions. The other velocity components appear to be set by global properties of the cloud, explaining the lack of variation. The front and flow velocities are determined by the large-scale velocity gradients within which the filament network is embedded, and the low value of \vrad is because on scales corresponding to the width of the sub-filaments, we expect the velocity dispersion to be subsonic  \citep{Heyer04}.

\begin{table}
	\centering
		\begin{tabular}{l c c c c}
   	         \hline
	         \hline
		  Simulation &  \vfront & $\vert$ \vflow $\vert$ &  \vrad & \vrot \\
		   & [\kms] & [\kms] & [\kms] &[\kms] \\
	         \hline                  
	         S1F1T238 & 0.74  & 0.38  & -0.100  & -0.030 \\
	         S1F1T260 & 0.70  & 0.33  & -0.089  & 0.008 \\
	         S1F1T281 & 0.71  & 0.30  & -0.102  & 0.010 \\
	         S1F1T303 & 0.75  & 0.36  & -0.128  & -0.005\\
	         S1F1T325 & 0.89  & 0.46  & -0.119  & 0.014 \\
		 \hline
		 S2F1T346 & 0.82  & 0.33  & -0.122 & -0.012 \\
		 S2F1T368 & 0.78  & 0.31 & -0.132 & -0.019 \\
		 S2F1T390 & 0.82  & 0.36 & -0.118 & 0.014 \\
		 S2F1T411 & 0.68  & 0.41 & -0.086 & 0.017 \\
		 S2F1T433 & 0.63  & 0.35 & -0.083 & 0.005 \\
		 \hline
	         \hline
		\end{tabular}
	
	\caption{ The evolution of the mean velocity components as the simulations evolve. The mean absolute magnitude of \vflow is shown so that it can be fairly compared to \vfront.  
	\label{evo_vel}}
\end{table}

\section{Discussion}\label{discussion}

\subsection{Comparison to observations}\label{obs_comp}

So far we have concentrated on a purely theoretical analysis of the sub-filament velocities, but it is also important to determine whether our results are consistent with observations. We will do this more fully in future work where we plan to do a detailed comparison of the derived filament velocities in the observational plane. However, this paper would be incomplete if we did not show that our sub-filaments actually resemble real position-position-velocity structures seen in molecular clouds.

Our approach is largely motivated by the observations of \cite{Hacar13} who show that the Taurus filament exhibited multiple narrow line components when observed in C$^{18}$O emission. It is this finding that allows them to deduce the existence of velocity coherent sub-filaments or `fibres' within the clouds. 

We have shown that hydrodynamic turbulent clouds contain a network of sub-filaments in agreement with observations, but we have not yet proven that this leads to the observed emission profiles. To test this hypothesis, we perform radiative transfer calculations of S1F1T281 and S2F1T368 using the \radmc code\footnote{This code is publicly available at the website www.ita.uni-heidelberg.de/$\sim$dullemond/software/radmc-3d/}.  \radmc models a variety of radiative processes including dust thermal emission and absorption, dust scattering and atomic and molecular line emission. It is this latter capability which is of key use here. We use our {\sc arepo} simulations as input for the density, temperature, velocities, and CO abundance, and then use \radmc to calculate the level populations and perform a ray-trace.

As C$^{18}$O is not necessarily in local thermodynamic equilibrium (LTE) throughout our filaments, we use the large velocity gradient (LVG) approximation \citep{Sobolev57}, which assumes that gas which is widely separated in position is also widely separated in velocity space, meaning that only local radiation will affect the level populations.\footnote{For a full description of how the LVG approximation has been implemented in \radmc, see \citet{Shetty11}.} As \radmc cannot work directly with the unstructured Voronoi mesh used within {\sc arepo}, we first map the simulation data onto a $400^3$ Cartesian grid with a pixel size of 0.008~pc before producing our synthetic emission maps. The C$^{18}$O isotope is not tracked directly in the chemical model, and so we assume that it has an abundance that is 557 times smaller than that of $^{12}$CO \citep{Wilson99}. To make the final image and spectra we then smooth the image with a gaussian beam. To test the effect of beam size we use three different beams with full width half maxima of 2, 4, and 8 pixels respectively. At the distance of Taurus (140 pc), these would correspond to beams of size $\sim 0.4'$, $0.8'$ and $1.6'$.

We do not adopt a preferred distance for our model, but instead calculate the intensity per pixel in physical units, not the angular size. This can be converted to a pixel scale in degrees once a choice is made as to the distance of the object. Since we are not modelling a specific region here we use the more general case, so that the reader can scale the values to the distance of their preferred object. In order to make the lines more easily comparable to brightness temperature, we convert our intensities into Kelvin by multiplying by the factor $\lambda^{2}/2k$, where $\lambda$ is the wavelength of the line and $k$ is Boltzmann's constant.

\begin{figure*}
\begin{center}
\includegraphics[width=5in]{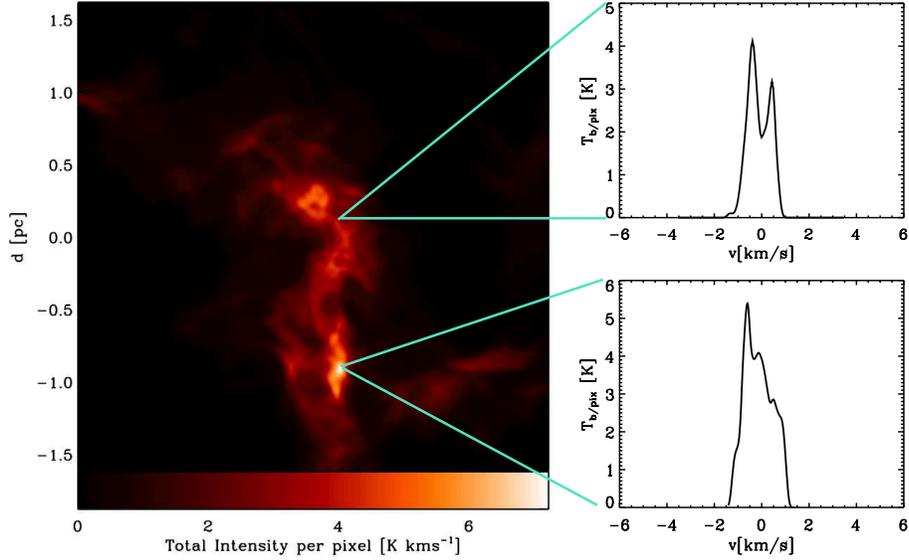}
\caption{Synthetic emission map for the C$^{18}$O (1-0) line, for simulation S1F1 \textbf{observed with a 4 pixel beam}. The panels on the right show the line profiles from the two highlighted regions. In agreement with observations there are multiple narrow velocity components seen in the line.}
\label{fig_c18o}
\end{center}
\end{figure*}

Figure \ref{fig_c18o} shows the resulting integrated emission map for the $J= 1 \rightarrow 0$ transition of C$^{18}$O, based on simulation S1F1T281 viewed in the $x$-$y$ plane (i.e.\ the observer's sightline is the $z$-axis in Fig \ref{columns}) with a 4 pixel FWHM beam. For illustrative purposes we show the line profiles at two points along the filament seen in the integrated emission. The line profiles show multiple velocity components in the C$^{18}$O~(1-0) line, in agreement with observations. Moreover, the line components are narrow due to the individual sub-filaments being coherent with subsonic velocities in the dense gas along the sub-filament spine as discussed in Section \ref{sect:rad_rot}. 

\begin{figure*}
\begin{center}
\begin{tabular}{c c}
\begin{overpic}[scale=0.4]{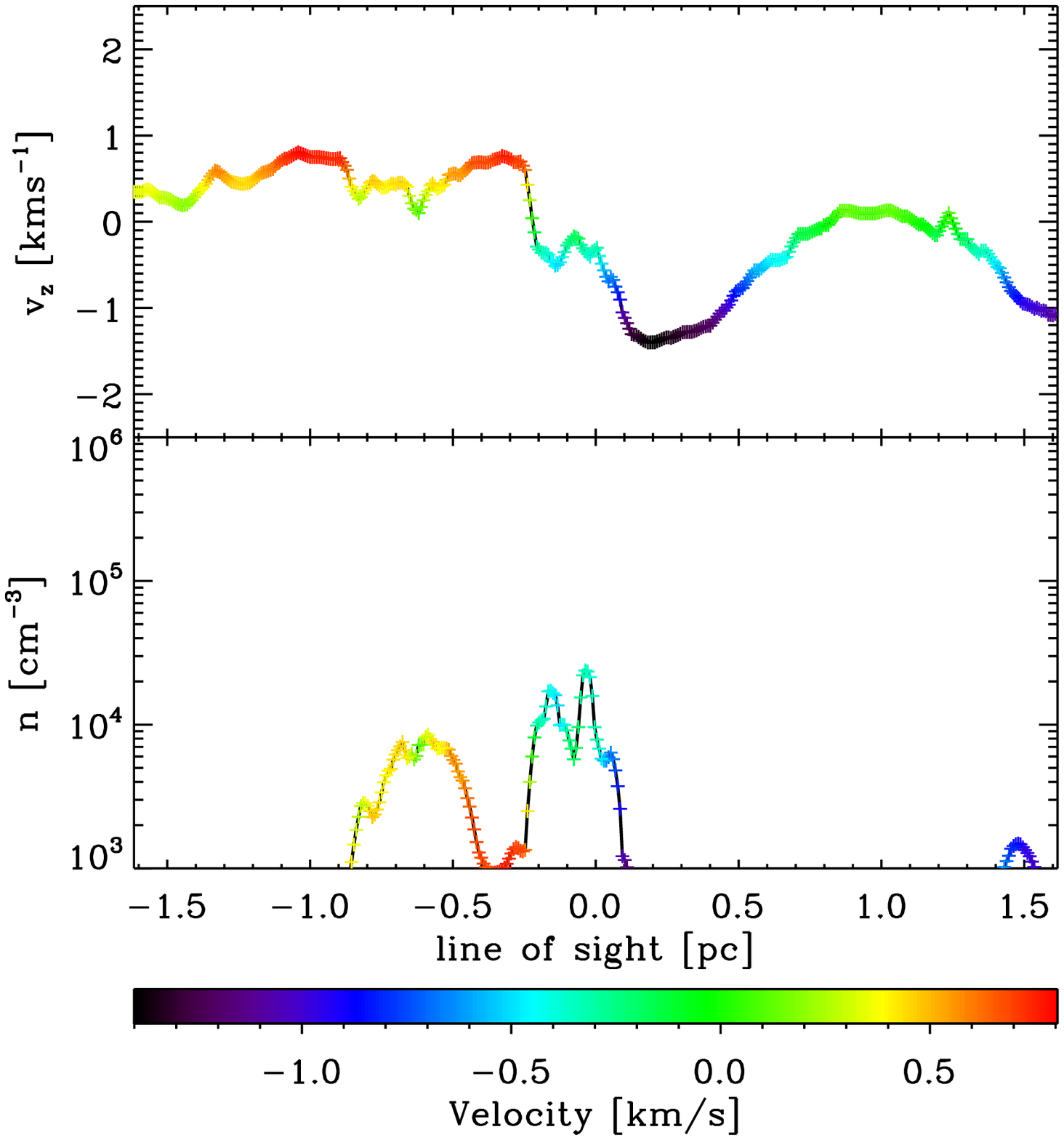}
\put (35,90) {\makebox(0,0){{\bf 2 line components}}}
\end{overpic}
\begin{overpic}[scale=0.4]{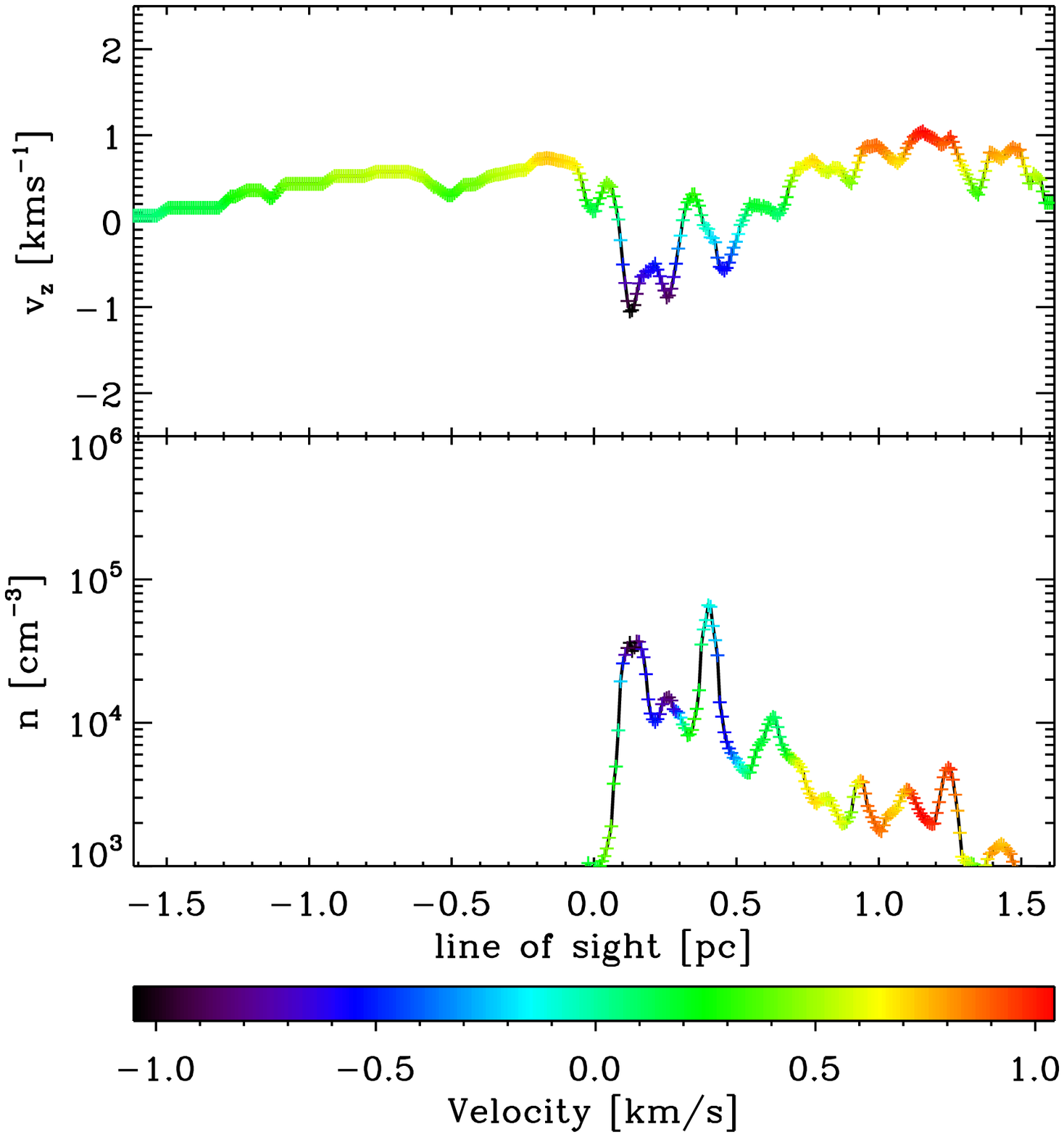}
\put (35,90) {\makebox(0,0){{\bf 3 line components}}}
\end{overpic}
\end{tabular}
\caption{The density and line-of-sight velocity along the beam for the two line profiles shown in Figure \ref{fig_c18o}. The colour scale also shows the line-of-sight velocity to make it easier to see the velocity at which the density field peaks. Typically the number of components in the line gives a lower limit to the number of over-densities in the beam.}
\label{sightline}
\end{center}
\end{figure*}

Figure \ref{sightline} explores the origin of these lines. The bottom panel shows the density along the observer's line of sight and the top shows the corresponding velocity, for the two line profiles in Figure \ref{fig_c18o}. The lines are colour-coded with their velocity to make it easy to see the velocity of the high density peaks along the sightline. We find there are at least as many density peaks as there are visible velocity components, meaning that the number of observed components gives a lower limit on the amount of substructure in a region similar to the one that we simulate.

We plan to carry out a full analysis of the peak velocities and line-widths that would be observed in our simulated clouds in future work where we will treat the problem in detail and include observational effects such as noise. However, in order to get an idea of how common multiple line components might be, we perform the following analysis. Using the synthetic position-position-velocity (PPV) cube generated by \radmc, we can obtain a line profile for each position in the image. When the line profile is bright (here defined as the peak emission being above 2$\,$K), we find the number of maxima in the profile. This brightness threshold is arbitrarily chosen to ensure that we are only including pixels which lie in the filament identified in the column density map.

\begin{table}
	\centering
		\begin{tabular}{l c c c c}
   	         \hline
	         \hline
		  Filament &  Transition & Beam & 2 maxima & More than 2  \\
		   & & [pix] & [\%] & [\%]\\
	         \hline                  
	         S1F1T281  & 1-0  & 2 & 34.12 & 11.18\\
	         S1F1T281  & 1-0  & 4 & 32.82 & 7.35 \\
	         S1F1T281  & 1-0  & 8 & 28.77 & 3.80 \\
	         \hline
	         S1F1T281 &  2-1  & 2 & 31.90 & 8.13\\
	         S1F1T281 &  2-1  & 4 & 28.16 & 4.70 \\
	         S1F1T281 &  2-1  & 8 & 21.41 & 1.18\\
	         \hline
	         \hline
	         S2F1T368 & 1-0  & 2 & 25.29 & 2.59 \\
	         S2F1T368 & 1-0  & 4 & 19.63 & 1.43 \\
	         S2F1T368 & 1-0  & 8 & 12.72 & 0.38 \\
	         \hline
	         S2F1T368 &  2-1 & 2 & 21.06 & 2.14 \\
	         S2F1T368 &  2-1 & 4 & 16.36 & 0.95 \\
	         S2F1T368 &  2-1 & 8 & 9.80 & 0.41\\
		 \hline
	         \hline
		\end{tabular}
	\caption{ The percentage of C$^{18}$O line profiles in the simulated PPV cube with a peak brightness greater than 2$\,$K that have multiple maxima.
	% should I add length to this table.
	\label{ncomponents}}
\end{table}

Table \ref{ncomponents} shows the percentage of bright pixels with multiple maxima in their line profiles for different simulations, line transitions and beam sizes. Some general trends emerge. The percentage of lines with multiple maxima varies between the simulated clouds, which is to be expected as they have different filament networks. There are more lines with multiple components in the lower energy $J= 1 \rightarrow 0$ line than the higher energy $J= 2 \rightarrow 1$ line. This is because some of the sub-critical sub-filaments have densities much lower than the $J= 2 \rightarrow 1$ critical density, and so the lines from these sub-filaments are faint. The number of line components also decreases as the beam size increases because the lines become broader and are more likely to be blended together. Regardless of this, the presence of multiple maxima in the emission lines is a ubiquitous feature of the synthetic observations.

It is not only in the predicted line emission that the filaments are in agreement with observations. \cite{Tafalla15} showed that core formation was only occurring in a subset of `fertile fibres' and that many of their observed fibres were not forming stars. When a fibre did form a core, it tended to form as part of a chain. In the simulations of collapsing turbulent clouds we observe similar behaviour. Most of the sub-filaments are sub-critical, but when they do reach criticality multiple sink particles (representing collapsing cores) tend to form, in agreement with the predictions of \cite{Inutsuka97}. Only very few sub-filaments become super-critical within  $10^6\,$ years of their formation. Most others need considerably longer to accrete sufficient mass -- if they manage at all. This fact may help to account for the low efficiency of star formation in molecular clouds.
%%% RSK: Rowan, please check the number 10^6 yr. Also: Should we mention that we have no feedback, which would make things even slower?

\subsection{Fray and gather}
From their observations \cite{Tafalla15} proposed a `fray and fragment' scenario where a filamentary density enhancement was first formed by two gas flows colliding, which then `frayed' into sub-filaments, and then fragmented into cores. In our simulations we have the advantage of being able to follow the time evolution of the gas. In the collapsing turbulent clouds the `fray' stage of the process where sub-filaments are formed occurs \textit{first}. This can be seen from the fact that the number of sub-filaments can either increase or decrease as the filament evolves (see Table \ref{evo-Nfil}), and that even in diffuse gas such as S2F2 there are many sub-filaments. Dense gas has this morphological appearance right from the beginning. These sub-filaments are then gathered together by large-scale motions within the cloud. Initially these are the large-scale turbulent modes, which have bulk velocities on the scale of the box length. However, as the turbulence decays the cloud becomes more and more gravitationally unstable which also gathers the sub-filaments together. This process was illustrated in \citeauthor{Smith14b}~(2014b) when we discussed the evolution of S2F1.

The formation of sub-filaments can also be thought of in terms of the turbulent cascade \citep{Kolmogorov41,Boldyrev02b}. \cite{Schmidt08} analysed high-resolution turbulent boxes with various forcing scales and found that the most dissipative structures are intermediate between filamentary and sheet-like structures. We propose that the sub-filaments are formed from the turbulent dissipation of turbulent modes with short wavelengths and that the filaments seen in column density are formed by long wavelength turbulent modes gathering the gas together. 

%Note: observationally hard to test- would have to look in more diffuse tracers.

\subsection{Caveats and future work}
This work has so far neglected the effects of magnetic fields. This was a deliberate choice in order to understand the velocity field arising from turbulence and gravity alone. \citet{Kirk15} studied the density profiles of filaments formed in simulations of MHD turbulence and found that magnetic fields provided additional pressure support which broadened the column density profiles and reduced fragmentation compared to the hydrodynamic case. \cite{Hennebelle13a} studied MHD simulations of non-self-gravitating filaments and suggested that ion-neutral friction may play a key role in determining the width of filaments. However at present there is no investigation of possible differences between filament velocities and sub-structure in the hydrodynamical and magnetohydrodynamical cases. This is something that we hope to address in future work.

We have also mainly confined ourselves to a theoretical analysis, with the exception of Section \ref{obs_comp} where we tested whether the simulations were compatible with observations. In future work we plan to carry out a fuller analysis, where we will test how synthetic observations of these clouds correspond to the simulations and how we may best derive the physical properties of the clouds from the observations. In previous work on filaments, we showed that star-forming cores embedded within filaments would have very different observational properties from those expected for isolated spherical cores. \cite{Smith12a} showed that the turbulent velocities in the diffuse envelopes of the filaments could  obscure the blue asymmetric line profile signature expected from a collapsing core. In \cite{Smith13a} we demonstrated that for massive protostars forming at the centre of a collapsing filamentary hub, the optically thick line profiles lacked self-absorption features and there were multiple optically thin line components in dense gas tracers.  In simple spherical collapse models one would expect strong self-absorption and a single narrow optically thin component at the line centre. We expect that mock observations of the sub-filament networks simulated in this paper will once again reveal line emission that could not have been predicted from simple analytical models of isolated star formation.

\section{Conclusions}\label{conclusions}
In this paper we have used high resolution \arepo simulations to show that turbulence in collapsing clouds naturally leads to the formation of a network of sub-filaments in agreement with recent observations of Taurus. Our main conclusions are as follows.

\begin{enumerate}
\item Filaments seen in column density maps actually consist of a network of sub-filaments when viewed in 3D.
\item The filaments are dynamical features that move according to the large-scale turbulent velocity field in the molecular cloud. These bulk velocities are predominantly perpendicular to the long axis of the filament.
\item There are flows of mass along the spine of the sub-filaments of the order of tens of solar masses per Myr. These can compress the sub-filament, increasing its mass and making it more likely to fragment. Alternatively the sub-filament may be sheared apart by turbulence, or form an accretion flow which feeds mass into a nearby star-forming region.
\item At the dense spine of the filament, where the density profile is flat, the gas velocities are both subsonic and coherent. The radial velocities are of order $-0.1 \: {\rm km \: s^{-1}}$ directed inwards, and there is no ordered rotation about the sub-filament spines.
\item Most sub-filaments are sub-critical according to the formalism of \cite{Inutsuka97} and do not fragment into cores. In the sub-filaments which are super-critical, multiple cores are forming, in agreement with observations.
\item Sub-filaments are a general feature of turbulent molecular clouds. They appear at all times and for all gas densities. 
\item The mass in sub-filaments and the number of super-critical sub-filaments increases as the molecular clouds evolve. In our simulated clouds the mass in sub-filament spines increased at a rate of $215 \: {\rm M_{\odot} \: Myr^{-1}}$ in S1F1, and by $97 \: {\rm M_{\odot} \: Myr^{-1}}$ in S2F1.
\item As the network of sub-filaments overlap along a typical observed sightline, this results in multiple narrow line components appearing in simulated C$^{18}$O line emission, in agreement with the observations of \cite{Hacar13}.
\item We propose that the formation of filaments and sub-filaments is a natural consequence of a turbulent cascade. Sub-filaments are formed by the high wavenumber, short wavelength modes. These are then gathered together by large-scale, low wavenumber modes and gravitational collapse to form the extended filaments observed in column density maps.
\end{enumerate}

\section*{Acknowledgements}
We would like to thank Adam Avison,  Doris Arzoumanian, Paul Clark, Alvaro Hacar, Patrick Hennebelle, Helen Kirk, Evangelia Ntormousi, Nicola Schneider, S\"{u}meyye Suri, Mario Tafalla, and Richard Wunsch for many useful discussions about filaments. RJS gratefully acknowledges support from the RAS through their Norman Lockyer Fellowship, and local support by the UK ARC node. SCOG and RSK acknowledge support from the Deutsche Forschungsgemeinschaft (DFG) via the Collaborative Research Centre SFB 881 {\em The Milky Way System} (sub-projects B1, B2 and B8) and the Priority Programme SPP 1573 {\em Physics of the ISM}. In addition, they also acknowledge support from the European Research Council under the European Community's Seventh Framework Programme (FP7/2007-2013) via the ERC Advanced Grant {\em STARLIGHT} (project number 339177). GAF acknowledges the support from the STFC consolidated grant ST/L000768/1. The simulations in this paper carried out on the MilkyWay cluster, which is supported by the DFG via SFB 881 (sub-project Z2) and SPP 1573.

\bibliography{./Bibliography}

\label{lastpage}

\end{document}